\documentclass[11pt]{article}

\newcommand{\blind}{0}

\makeatletter
\renewcommand\section{\@startsection {section}{1}{\z@}%
                                   {-3.5ex \@plus -1ex \@minus -.2ex}%
                                   {2.3ex \@plus.2ex}%
                                   {\normalfont\sffamily\fontsize{16}{19}\bfseries}}
\renewcommand\subsection{\@startsection{subsection}{2}{\z@}%
                                     {-3.25ex\@plus -1ex \@minus -.2ex}%
                                     {1.5ex \@plus .2ex}%
                                     {\normalfont\sffamily\fontsize{14}{17}\bfseries}}
\renewcommand\subsubsection{\@startsection{subsubsection}{3}{\z@}%
                                    {-3.25ex\@plus -1ex \@minus -.2ex}%
                                     {1.5ex \@plus .2ex}%
                                     {\normalfont\normalsize\sffamily\fontsize{14}{17}\selectfont}}
\makeatother

\usepackage{hyperref}

\usepackage{amsmath}
\usepackage{amsfonts,amssymb,amsthm}
\usepackage{graphicx}
\usepackage{enumerate}
\usepackage{xcolor}
\usepackage{natbib}
\usepackage{url}
\usepackage{booktabs}
\usepackage{algorithm}
\usepackage{algpseudocode}
\usepackage{makecell}
\usepackage{rotating}
\usepackage{multirow}

\usepackage{mathtools}
\usepackage{physics}
\usepackage{cleveref}
\usepackage[margin=1in]{geometry}
\usepackage{setspace}
    
    \crefname{assumption}{assumption}{assumptions}
\usepackage{bbm}

\usepackage{threeparttable}
\usepackage{marvosym}

\theoremstyle{definition}

\newcommand{\bLambda}{\boldsymbol{\Lambda}}
\newcommand{\blambda}{\boldsymbol{\lambda}}

\newcommand{\bk}{\boldsymbol{k}}

\newcommand{\bb}{\boldsymbol{b}}
\newcommand{\bs}{\boldsymbol{s}}

\def\calB{\mathcal{B}}
\def\calP{\mathcal{P}}
\def\calN{\mathcal{N}}

\DeclareMathOperator{\E}{\mathbb{E}}
\def\NatInt{\mathbb{N}}
\begin{document}

	
	\if0\blind
	{
		\title{\bf Two-Stage Staffing with Random Arrival Rate: A Simulation Optimization Approach}
		\author{Xu Yang$^\textrm{a}$, Haihui Shen$^{\textrm{b},\textrm{c},}$\thanks{Corresponding Author {\large\Letter} \href{mailto:shenhaihui@sjtu.edu.cn}{shenhaihui@sjtu.edu.cn}}\\
        {\small$^\textrm{a}$Antai College of Economics and Management, Shanghai Jiao Tong University, Shanghai, 200030, China}\\
		{\small$^\textrm{b}$Sino-US Global Logistics Institute, Antai College of Economics and Management,}\\
        {\small Shanghai Jiao Tong University, Shanghai, 200030, China}\\
        {\small$^\textrm{c}$Data-Driven Management Decision Making Lab, Shanghai Jiao Tong University,}\\
        {\small Shanghai, 200030, China}}

		\date{November, 2026}

        \vspace*{-30pt}
        {\let\newpage\relax\maketitle}
	} \fi
	
	\if1\blind
	{
		\title{\bf Two-Stage Staffing with Random Arrival Rate: A Simulation Optimization Approach}
		\author{Author information is purposely removed for double-blind review}

        \date{October, 2026}

        \maketitle

	} \fi
	\bigskip
	
\begin{abstract}
In service systems, the arrival rate in the same time period and day of the week exhibits randomness, yet can be more accurately estimated or forecast right before that period, which motivates a two-stage staffing scheme with advance base staffing and pre-shift surge staffing.
Existing approaches usually impose strong assumptions on the arrival process and service systems' operational logic and other further approximations (i.e., stylized models) to derive explicit staffing policies.
This paper proposes a simulation optimization approach (SOA) for more general arrival processes and service systems by effectively addressing some intertwined challenges.
By recasting the original formulation into computationally tractable structures and designing an efficient coordinate-search algorithm,
it constructs a two-stage staffing lookup table offline, which allows managers to make real-time surge staffing decisions without any further simulation.
Comprehensive numerical experiments demonstrate that the proposed SOA has significantly better staffing performance and substantially broader applicability compared to existing approaches.
Even under the stylized models,
the cost savings delivered by our SOA compared with existing approaches range from the low tens to several tens of percent, depending on different problem parameters;
the advantage further amplifies when the actual problem deviates from the stylized models.
In a nurse-staffing case study grounded in real-world data, the proposed SOA delivers approximately 8\% to 13\% cost savings over a state-of-the-art benchmark.

\end{abstract}

\noindent%


\doublespacing

\section{Introduction}\label{s:intro}

In many service systems, staffing decisions play a central role in matching service capacity to demand. For example, in emergency departments (EDs), inadequate nurse staffing is associated with longer patient lengths of stay and higher leaving without being seen (LWBS) rates \citep{drennanAssociationNurseStaffing2024}.
Similarly, in call centers, insufficient agent staffing can increase customer waiting time and abandonment rates \citep{garnettDesigningCallCenter2002}.
Conversely, excess staffing leaves capacity underutilized, with high labor costs accounting for a large proportion of operational expenses.
Therefore, balancing staffing levels (and cost) against quality of service (QoS) is a key issue in service systems.

However, making effective staffing decisions is challenging, largely because the service demand is often uncertain and varies over time.
For example, although the arrivals in healthcare systems or call centers may exhibit certain within-day and within-week patterns, accurately predicting the number of arrivals or timing of arrivals in a given time period is very challenging \citep{greenCopingTimevaryingDemand2007}.
Empirical studies show that the number of arrivals in the same time period and day of the week across multiple weeks often exhibits overdispersion, i.e., the variance is significantly greater than the mean, which suggests that a standard non-homogeneous Poisson process is often not capable of characterizing the real arrival process \citep{kimAreCallCenter2014}.
Many studies, e.g., staffing in healthcare systems \citep{huPredictionDrivenSurgePlanning2025} and  call centers \citep{whittStaffingCallCenter2006,kocagaStaffingCallCenters2015}, tried to model the arrival process as a doubly stochastic Poisson process (DSPP), where the arrival rate of the Poisson process during some time period is a random variable.

In practice, staffing decisions in service systems often need to be made weeks in advance.
But if the arrival rate during some time period itself is random, the staffing level planned based on the mean (i.e., expectation) of the arrival rate in advance must mismatch the actual arrival rate realized when that day comes \citep{bassambooCapacitySizingParameter2010,kocagaStaffingCallCenters2015}.
On the other hand, research indicates that short-term forecasting of arrival rates can incorporate a wealth of real-time information, thereby achieving superior predictive accuracy compared to long-term forecasting \citep{whittForecastingArrivalsOccupancy2019,huUseRealTimeInformation2023}.
This naturally motivates a two-stage staffing scheme, where the staffing plan made in advance can be adjusted once the arrival rate is more accurately estimated or forecast, via overtime or temporary staff, albeit at higher unit cost.
For example,
\citet{kocagaStaffingCallCenters2015} consider call centers where managers first determine in-house staffing and then, once the arrival rate is known, decide whether to outsource incoming calls based on system congestion.
\citet{kimTwoStageStochasticInteger2015} use two-stage stochastic programming to optimize both nurse staffing and scheduling, with initial plans adjusted in response to updated demand forecasts.
\citet{taSimulationBasedDecomposition2021} study staffing in call centers, where staffing level is first determined according to the distribution of arrival rate and adjusted in the second stage with updated forecasts, to minimize the staff cost while maintaining certain QoS.

A more recent and notable work is \citet{huPredictionDrivenSurgePlanning2025}, which considers the two-stage nurse staffing in EDs.
Patient arrivals are assumed to follow a DSPP, and the ED is assumed to operate as an $M/M/k + M$ queue given the realized arrival rate and specific QoS is considered.
Under this setting, they derive an explicit two-stage quality-and-efficiency driven (QED) staffing policy for the single-shift case based on some analytical approximations, with base staffing planned in advance and surge staffing determined once the arrival rate of that shift is known.
They prove that the two-stage QED policy is near-optimal in an asymptotic sense for the single-shift case, and suggest some heuristic adjustments for the multi-shift case.
Numerical experiments including a case study with real data demonstrate that the policy has satisfactory performance.
Nevertheless, several limitations remain.
First, the assumptions of the stylized models for the arrival process and service system limit its application in a more realistic or broader setting.
Second, the asymptotic optimality property does not necessarily ensure optimality in reality, when the service system operates in a transient state due to the time-varying arrivals and resources.
Third, the adjustments for the multi-shift case require coefficient calibration, making its performance potentially problem instance-specific.

In this paper, we propose a simulation optimization approach (SOA) framework for general two-stage staffing problems with random arrival rates, including, but not limited to, the
specific nurse staffing in EDs considered in \citet{huPredictionDrivenSurgePlanning2025}.
It is motivated by the wide application of simulation optimization in operations management \citep{amaranSimulationOptimizationReview2016}.
For staffing problems, \citet{feldmanStaffingTimeVaryingQueues2008} propose a simulation-based iterative staffing algorithm for time-varying queues to achieve time-stable performance.
In healthcare, simulation optimization has been widely used for resource planning,
including ED resource allocation \citep{ahmedSimulationOptimizationEmergency2009},
long-term care capacity planning \citep{zhangSimulationOptimizationApproach2012},
and ED physician staffing and scheduling \citep{wangSimulationbasedEmergencyDepartment2025};
see \citet{yousefiSimulationbasedOptimizationMethods2020} and \citet{wangSimulationOptimizationHealthcare2023} for a review.
These studies show the value of simulation optimization in evaluating and improving staffing and resource allocation decisions,
when complicated system dynamics make analytical solutions difficult to obtain.
To the best of our knowledge, however, the literature remains silent on directly applying simulation optimization to the two-stage staffing problem with random arrival rates.

The most immediate benefit of the SOA is that it allows general arrival processes with random rates and service systems that operate under arbitrary yet well-defined operational logic, not limited to the stylized models.
However, applying simulation optimization to such a two-stage staffing problem is far from trivial and inherently accompanied by critical challenges.
First, as will be elaborated in the sequel, the original formulation of the two-stage staffing problem for the general multi-shift setting constitutes a highly complex dynamic program.
Even leaving aside the time required to estimate the objective function via simulation, a direct solution to this dynamic programming problem is practically
intractable.
Second, the simulation time remains a non-negligible factor, despite utilizing robust computational resources and efficient techniques.
Third, even if solving the simulation optimization is computationally viable within a reasonable time, the second-stage surge staffing decision must be made rapidly after the arrival rate becomes known.
Such a strict (near-)real-time requirement is inherently beyond the reach of instantaneous simulation optimization.
These intertwined challenges are effectively addressed in this paper.

The main contributions of this paper can be highlighted as follows.
To the best of our knowledge, this study represents the first attempt to leverage simulation optimization directly for the general two-stage staffing problem under random arrival rates, accomplished through our proposed SOA framework.
To deal with the computational intractability of the original dynamic program and the real-time requirement for surge staffing decisions, we recast the multi-shift problem via tractable approximation, solve the resulting simulation optimization efficiently through tailored coordinate-search algorithms, and shift the entire computational burden offline.
Specifically, offline-generated arrival-rate paths are utilized to pre-compute the surge staffing decisions across arrival-rate intervals, yielding a two-stage staffing lookup table.
Once the true arrival rate is revealed prior to a shift, managers can instantaneously retrieve the surge staffing decision in real time without any further simulation, just like using closed-form formulas.
Comprehensive numerical experiments demonstrate that the proposed SOA has significantly better staffing performance and substantially broader applicability compared to existing approaches.
Moreover, our study reveals when the complexity of two-stage staffing will offer meaningful gains over the simple one-stage schedule.

The rest of the paper is organized as follows.
In \Cref{s:problem} we present the original formulation of the two-stage staffing problem in detail and discuss the main difficulty for solving it.
In \Cref{sec:SOA}, the SOA for the single-shift case and general multi-shift case is presented, respectively, followed by the comparison with existing policies.
\Cref{s:experiments} includes comprehensive numerical experiments, demonstrating the performance and superiority of the SOA,
and \Cref{s:conclusion} briefly concludes the paper.
Some additional numerical results are collected in the Appendix.

\section{Problem Formulation}\label{s:problem}

We consider a general two-stage staffing problem in a service system.
Suppose there are totally $I$ consecutive shifts, and the initial state of the first shift is given as $\pi_0$.
For a given shift $i$, $i=1,\ldots,I$, the demand arrives following a certain counting process, 
where $\lambda_i$ is a constant arrival rate during shift $i$.
In principle, the service system can operate under arbitrary yet well-defined operational logic.
Usually, one may consider a multi-server queuing system, possibly with some specific features, including queue discipline (first-come-first-served, shortest-processing-time-first, etc.), queue behavior (balking, reneging, etc.), and so on.
However, when deciding the base staffing level (i.e., number of servers) $\bb \coloneqq (b_1,\ldots,b_I) \in \NatInt^I$ some time in advance, the actual arrival rates are unknown (or unrealized) yet.

The arrival rates $\lambda_i$, $i=1,\ldots,I$, are modeled as random variables $\Lambda_i$ with marginal distribution $F_i$, $i=1,\ldots,I$.
We assume that their joint distribution $F$ is known and allow the arrival rates to be correlated across shifts.
Following \citet{huPredictionDrivenSurgePlanning2025}, this paper simplifies the environment by assuming that the actual arrival rates are perfectly revealed prior to each shift.
We recognize, however, that in practice they must be forecast with real-time information in some way,
and thus are inevitably subject to errors, despite being substantially more accurate than simple expected values \citep{huUseRealTimeInformation2023}.
Incorporating such forecasting errors into the model is a valuable topic for future research, but is beyond the scope of this study.
For ease of exposition, in the sequel we will simply state that \(\lambda _{i}\) is \textit{known} or \textit{observed} prior to the beginning of shift \(i\).

Suppose the base staffing level $\bb$ has been decided.
Prior to the beginning of shift $1$, the actual arrival rate $\lambda_1$ is known and the surge staffing level is decided to be $s_1 \in \NatInt$, where one surge staff is the same as one base staff in terms of service rate and quality, but costs more.
When shift $1$ begins, the total staffing level is $k_1 = b_1 + s_1$.
It reflects two operational issues in practice.
One is that it is difficult to downsize scheduled base staff at short notice, and the other is that it costs more to add staff at short notice.
Then prior to the beginning of shift $2$, the actual arrival rate $\lambda_2$ is known, together with the initial state of shift $2$ (i.e., the terminating state of shift $1$) that is denoted as $\pi_1$,
and surge staffing level $s_2\in \NatInt$ needs to be determined.
It follows likewise for $i=3,\ldots,I$.
To summarize, the base staffing level $\bb$ is determined and fixed in advance with knowledge of the joint distribution $F$,
the surge staffing level $\bs \coloneqq (s_1,\ldots,s_I )\in \NatInt^I$ is decided sequentially, where each $s_i$ is determined upon knowing $\lambda_i$ and $\pi_{i-1}$ prior to the beginning of shift $i$,
and the total staffing level $k_i = b_i + s_i$ is actually set in shift $i$, for $i=1,\ldots,I$.

The objective is to minimize the sum of the cost of staffing and the expected penalty related to the QoS of the service system.
Denote $f_i\qty (k_i,\lambda_i,\pi_{i-1})$ as the expected penalty related to the QoS of the service system with initial state $\pi_{i-1}$ when the actual staffing level is $k_i$ and the arrival rate during the shift is $\lambda_i$.
Here the expectation is taken with respect to the randomness in the queuing system in shift $i$.
And denote $c_i^{(b)}$ and $c_i^{(s)}$ as the unit cost of base and surge staff, respectively, for $i=1,\ldots,I$.
Note that $f_i(\cdot)\geq 0$ and $c_i^{(s)} > c_i^{(b)} > 0$.

\subsection{Simplified Formulation for Single Shift}

Let us first consider the simplified case where $I=1$, or treat the service system as if the $I$ ($I>1$) shifts were independent and the initial state of each shift were given.
In the latter case, the optimization problem just degenerates to $I$ isolated single-shift problems with a known initial state.
In either case, we only need to focus on one shift, say $i$.
Upon knowing $\lambda_i$, the second-stage optimization problem is defined as
\begin{equation}
	\label{eq:original_model_second_stage}
	z\qty (b_i, \lambda_i,\pi_{i-1}) \coloneqq \min_{s_i \in \NatInt} \qty{ c_i^{(s)} s_i + f_i \qty (b_i+s_i,\lambda_i,\pi_{i-1}) },
\end{equation}
and the first-stage optimization problem is defined as
\begin{equation}
	\label{eq:original_model_first_stage}
b_i^* \coloneqq  \arg\min_{b_i \in \NatInt}  \qty{ c_i^{(b)} b_i + \E_{\Lambda_i} \qty [z\qty (b_i, \Lambda_i,\pi_{i-1})] },
\end{equation}
where the expectation is taken with respect to the random arrival rate $\Lambda_i \sim F_i$.
Given the optimal base staffing level $b_i^*$, the optimal surge staffing level upon knowing $\lambda_i$ is
\begin{equation}\label{eq:original_model_surge_stage}
s_i^* \coloneqq  \arg\min_{s_i \in \NatInt} \qty{  c_i^{(s)} s_i + f_i \qty (b_i^*+s_i,\lambda_i,\pi_{i-1}) }.
\end{equation}

\subsection{General Formulation for Multiple Shifts}

Unfortunately, $I$ shifts are usually dependent when $I>1$, as the unfinished customers in one shift usually carry over to the next shift in practical service systems.
That is, the initial state of shift $i$, $\pi_{i-1}$, is not given in advance, but a consequence of the arrival rates and staffing decisions in all previous shifts (subject to randomness in the queuing system).
Let $\Pi_{i-1}$ denote the random initial state of shift $i$.
Suppose the base staffing level $\bb\in \NatInt^I$ has been decided.
Let $Y_i$ denote the available information besides $\bb$ when deciding the surge staffing level for shift $i$,
and let $S_i$ denote the corresponding surge staffing decision (which is also a random variable determined by $Y_i$ and $\bb$).
Then we can see that
\begin{equation*}
\begin{split}
  Y_1 & \coloneqq (\pi_{0}, \Lambda_1),\\
  Y_{i+1} & \coloneqq (Y_i, S_i, \Pi_i, \Lambda_{i+1}), \quad i=1,\ldots, I-1.
\end{split}
\end{equation*}
Note that $\Pi_i$ is actually determined by $\Pi_{i-1}$, $\Lambda_{i}$, and $b_i+S_i$, together with the randomness in the queuing system in shift $i$.

Suppose $Y_i$ is realized as $y_i$, where $\Pi_{i-1}$ is realized as $\pi_{i-1}$ and $\Lambda_{i}$ is realized as $\lambda_i$.
Let $V_i(y_i; \bb)$ denote the expected remaining cumulative cost from shift $i$ until shift $I$ if the surge staffing levels for all $(I-i+1)$ shifts are sequentially and optimally decided.
Let $V_{I+1}(\cdot; \bb) = 0$, and then for $i = 1, \ldots, I$, we have the recursive relationship:
\begin{equation*}
V_i(y_i; \bb) = \min_{s_i \in \NatInt} \qty{ c_i^{(s)} s_i + f_i \qty (b_i+s_i,\lambda_i,\pi_{i-1}) + \E_{\Pi_i, \Lambda_{i+1}}\qty[V_{i+1}(Y_{i+1}; \bb) \, \big| \, Y_i=y_i, S_i = s_i ] },
\end{equation*}
where the expectation is taken with respect to the random initial state $\Pi_i$ and random arrival rate $\Lambda_{i+1}$ in shift $(i+1)$.
Then the first-stage optimization problem for deciding the optimal base staffing level $\bb$ is
\begin{equation} \label{eq:original_multi_model_first_stage}
\bb^* \coloneqq \arg\min_{\bb \in \NatInt^I}  \qty{ \sum_{i=1}^I c_i^{(b)} b_i   + \E_{\Lambda_1} \qty [V_1(Y_1; \bb)] },
\end{equation}
where the expectation is taken with respect to the random arrival rate $\Lambda_1$ in shift $1$.
Note that the optimization of surge staffing is still performed sequentially, as the actual arrival rates are revealed sequentially.
Upon knowing $y_i$ prior to the beginning of shift $i$, $i=1,\ldots,I$, the optimal surge staffing level for shift $i$ is
\begin{equation}\label{eq:original_multi_model_surge_decision}
s_i^* \coloneqq \arg\min_{s_i \in \NatInt} \qty{ c_i^{(s)} s_i + f_i \qty (b_i^*+s_i,\lambda_i,\pi_{i-1}) + \E_{\Pi_i, \Lambda_{i+1}}\qty[V_{i+1}(Y_{i+1}; \bb^*) \, \big| \, Y_i=y_i, S_i = s_i ] }.
\end{equation}

\subsection{The Main Difficulty}

The above optimization problems are mathematically accurate, but it is almost impossible to directly solve them.
First, consider the simplified formulation for the single-shift case.
The main difficulty lies in the intractability of $f_i\qty (k_i,\lambda_i,\pi_{i-1})$.
Over a finite shift horizon (e.g., $8$ or $12$ hours), transient dynamics and the initial state $\pi_{i-1}$ play an important role, so the system cannot be treated as operating in steady state.
Thus the expected penalty related to the QoS of the service system is difficult to derive, even if we assume some stylized queuing models with well-known steady-state distributions.
And in fact some specific queue features, such as queue discipline and queue behavior, need to be considered in the service system in order to be realistic.
It makes the mathematical derivation of $f_i(\cdot)$ even less likely.
Approximation and simulation-based estimation seem to be the only two possible methods.

When it comes to the general formulation for the multi-shift case, this formulation significantly magnifies the difficulty.
First, the system state will transit among consecutive shifts, which means the distribution of system states in different shifts should be analyzed jointly.
An exact evaluation is extremely difficult; even a simulation-based estimation is highly challenging, to say nothing of theoretical analysis or approximation.
Moreover, the defined expected remaining cumulative cost $V_i(y_i; \bb)$ involves both challenges of expectation estimation and iterative optimization.
Solving such a dynamic programming problem directly is practically intractable for moderate values of $I$, even if we can efficiently simulate the service system.

\citet{huPredictionDrivenSurgePlanning2025} take an analytical approximation approach to two-stage staffing.
Specifically, they impose stylized models for the arrival process and service system with specific QoS considered (which will be elaborated in detail later).
For the single-shift case, by further assuming the system operates in a steady state, they first derive a two-stage Newsvendor rule using a stochastic-fluid approximation and then refine it using a diffusion approximation to obtain the two-stage QED staffing rule.
For the general multi-shift case, some heuristic adjustments to the previous staffing rule are suggested to compensate for the transient-shift effect.
In this paper, we take a simulation optimization approach for much more general settings.
As previously noted and elaborated further below, applying simulation to this problem---particularly in the multi-shift case---is not straightforward, requiring an appropriate reformulation of the original problem.

\section{Simulation Optimization Approach for Two-Stage Staffing} \label{sec:SOA}

In this section, we first present the SOA for the single-shift case, then extend it to the full version for the more general multi-shift case.
We will also compare the SOA with existing approaches in terms of both problem setting and staffing policy.

\subsection{SOA for Single Shift}

Recall that $f_i\qty (k_i,\lambda_i,\pi_{i-1})$ is the expected penalty related to the QoS of the service system with initial state $\pi_{i-1}$ when the actual staffing level is $k_i$ and the arrival rate during the shift is $\lambda_i$.
Since the service system will usually not enter a steady state during one shift, $f_i(\cdot)$ should be estimated by running terminating simulation, where the time duration is the shift length.
Note that $\pi_{i-1}$ is given and fixed in the single-shift case.

To solve the optimization problems \eqref{eq:original_model_first_stage} and \eqref{eq:original_model_surge_stage}, we adopt the classical two-layer nested simulation.
In the outer layer, $M$ values of $\Lambda_i$ are sampled from the distribution of $\Lambda_i$, $F_i$, which are denoted as $\lambda_{im}$, $m=1,\ldots,M$.
In the inner layer, $f_i\qty (k_i,\lambda_i,\pi_{i-1})$ is estimated for given $k_i$ and $\lambda_i$.
Let $f_{ir}\qty (k_i,\lambda_i,\pi_{i-1})$ denote the calculated penalty in the $r$-th terminating simulation run, and totally $R$ independent simulation replications are conducted.
Then $f_i\qty (k_i,\lambda_i,\pi_{i-1})$ is estimated by
\begin{equation*}
\bar{f}_{iR}\qty (k_i,\lambda_i,\pi_{i-1}) \coloneqq \frac{1}{R}\sum_{r=1}^R f_{ir}\qty (k_i,\lambda_i,\pi_{i-1}).
\end{equation*}
To more efficiently compare $\bar{f}_{iR}\qty (k_i,\lambda_i,\pi_{i-1})$ for different values of $k_i$, one can use common random numbers (CRN) across different values of $k_i$ under the same $\lambda_i$.
Specifically, the simulations with different values of $k_i$ share the same set of exogenous random inputs, including random arrivals, as well as their service times and other personal characteristics.

Given the first-stage decision $b_i$, for each outer-layer realization $\lambda_{im}$, the second-stage optimization problem becomes
\begin{equation}
	\label{eq:simulated_model_second_stage}
\bar{z}_R \qty (b_i, \lambda_{im},\pi_{i-1}) \coloneqq \min_{s_i \in \NatInt} \qty{ c_i^{(s)} s_i + \bar{f}_{iR} \qty (b_i+s_i,\lambda_{im},\pi_{i-1}) }.
\end{equation}
So, the first-stage optimization problem for deciding the optimal base staffing level $b_i$ in SOA is
\begin{equation}
\label{eq:simulated_model_first_stage}
\bar{b}_i^* \coloneqq  \arg\min_{b_i \in \NatInt}  \qty{ c_i^{(b)} b_i + \frac{1}{M} \sum_{m=1}^M \bar{z}_R \qty (b_i, \lambda_{im},\pi_{i-1}) }.
\end{equation}
The optimal surge staffing level upon knowing $\lambda_i$ in SOA is
\begin{equation}\label{eq:simulated_model_surge_stage}
\bar{s}_i^* \coloneqq  \arg\min_{s_i \in \NatInt} \qty{  c_i^{(s)} s_i + \bar{f}_{iR} \qty (\bar{b}_i^*+s_i,\lambda_i,\pi_{i-1}) }.
\end{equation}

Since both Problems \eqref{eq:simulated_model_second_stage}-\eqref{eq:simulated_model_first_stage} are one-dimensional integer simulation optimization problems, we simply employ a hill-climbing search strategy.
That is, to solve $\min_x g(x)$ with currently found solution $x$, iteratively move to the neighbor ($x-1$ or $x+1$) with the smallest estimated function value, if it is lower than the estimated value of $g(x)$;
and terminate otherwise.
For each candidate base staffing level $b_i$, we use an inner hill-climbing search to find a local solution in the second-stage Problem \eqref{eq:simulated_model_second_stage} under each of the $M$ realizations of $\Lambda_i$, where each $\bar{f}_{iR}$ is computed from $R$ simulation replications of the service system.
With each estimated $\bar{z}_R \qty (b_i, \lambda_{im},\pi_{i-1})$, we can estimate the objective function in the first stage Problem \eqref{eq:simulated_model_first_stage} for each $b_i$.
Based on this, an outer hill-climbing search is used to solve $\bar{b}_i^*$.
In the implementation, to avoid redundant simulation, we cache the simulation estimates $\bar{f}_{iR}\qty (k_i,\lambda_i,\pi_{i-1})$ by $(k_i,\lambda_i)$.
CRN is used when evaluating $\bar{f}_{iR}\qty (k_i,\lambda_i,\pi_{i-1})$ for different values of $k_i$ under the same $\lambda_i$.
Moreover, for each $b_i$, the inner searches across the $M$ realizations of $\Lambda_i$ can be performed in parallel.

Solving $\bar{s}_i^*$ in \eqref{eq:simulated_model_surge_stage} is just similar to solving Problem \eqref{eq:simulated_model_second_stage} once with $b_i = \bar{b}_i^*$ and $\lambda_{im}=\lambda_i$.
So the computation delay may not significantly impede the near-real-time decision making upon knowing $\lambda_i$ prior to the beginning of shift $i$.
If strict real-time decision-making is required, one can adopt the so-called offline simulation online application idea \citep{hongOfflineSimulationOnline2019}.
That is to say, pre-compute $\bar{s}_i^*$ for several values of $\lambda_i$ before observing it (i.e., solve \eqref{eq:simulated_model_surge_stage} for different values of $\lambda_i$), and report the most applicable $\bar{s}_i^*$ upon observing $\lambda_i$.
We will illustrate this issue in detail and adopt a simple method in the multi-shift case, where the computation delay indeed constitutes a non-trivial barrier for (near-)real-time decision making.

\subsection{SOA for Multiple Shifts}

Note that directly solving the optimization problems \eqref{eq:original_multi_model_first_stage} and \eqref{eq:original_multi_model_surge_decision} is practically
intractable, because the computation burden of both simulation and optimization will be easily prohibitive for moderate values of $I$.
To tackle this challenge, an approximation of the original optimization is considered.

\subsubsection{The Approximated Model} \label{sec:approximated_model}
To calculate the optimal base staffing level $\bb$, consider a more tractable scenario, where all actual arrival rates are revealed simultaneously (instead of sequentially).
For given actual staffing level $\bk \coloneqq (k_1,\ldots, k_I)$ and arrival rate $\blambda \coloneqq (\lambda_1,\ldots, \lambda_I)$,
define
\begin{equation}\label{eq:g}
g(\bk,\blambda,\pi_0) \coloneqq \E_{\Pi_1,\ldots,\Pi_{I-1}} \qty [\sum_{i=1}^I f_i\qty (k_i,\lambda_i,\Pi_{i-1}) \ \Big| \ \Pi_0=\pi_0 ],
\end{equation}
which means the entire cumulative expected penalty related to the QoS of the service system.
Note that the expectation in \eqref{eq:g} is taken with respect to $(\Pi_1,\ldots,\Pi_{I-1})$, since they are random for given $\bk$ and $\blambda$ due to the randomness in the queuing system.
Then, the optimization problem for deciding the optimal base staffing level $\bb$ becomes
\begin{equation} \label{eq:approximate_multi_model_first_stage}
\tilde{\bb}^* \coloneqq \arg\min_{\bb \in \NatInt^I}  \qty{ \sum_{i=1}^I c_i^{(b)} b_i   + \E_{\bLambda} \qty [ \min_{\bs\in\NatInt^I} \qty{ \sum_{i=1}^I c_i^{(s)} s_i + g(\bb+\bs,\bLambda,\pi_0) } ] },
\end{equation}
where the expectation is taken with respect to the random arrival rate vector $\bLambda \coloneqq (\Lambda_1,\ldots,\Lambda_I)$.

Although the base staffing level is decided by treating the arrival rates as if they were revealed simultaneously,
the surge staffing level still needs to be decided sequentially.
Prior to the beginning of shift $i$, $\lambda_i$ and $\pi_{i-1}$ will be observed.
Strictly speaking, the optimal $s_i$ must depend on both $\lambda_i$ and $\pi_{i-1}$.
But $\pi_{i-1}$ could be a set with tremendous variables (which together define the state of a system), in the simulation viewpoint explicit dependence on $\pi_{i-1}$ means we have to discard all the previous simulated sample paths, and restart from $\pi_{i-1}$ to the final shift.
Such computation delay will make the (near-)real-time decision making impossible.
To reduce the computation burden, we use an approximation that avoids re-optimizing surge staffing for the observed initial state.
More specifically, let
\begin{equation*}
\bs(\tilde{\bb}^*,\blambda,\pi_0) \coloneqq \arg\min_{\bs\in\NatInt^I} \qty{ \sum_{i=1}^I c_i^{(s)} s_i + g(\tilde{\bb}^*+\bs,\blambda,\pi_0) },
\end{equation*}
and let $s_i(\tilde{\bb}^*,\blambda,\pi_0)$ denote the $i$-th element of $\bs(\tilde{\bb}^*,\blambda,\pi_0)$.
Prior to the beginning of shift $i$, we observe $\lambda_i$ and decide the surge staffing level as
\begin{equation}\label{eq:tilde_s_i}
\tilde{s}^*_i \coloneqq \textrm{round} \qty(\E \qty [s_i(\tilde{\bb}^*,\bLambda,\pi_0) \big| \Lambda_i = \lambda_i]  ).
\end{equation}

\subsubsection{Simulation-Based Solution} \label{sec:approximated_solution}

Similar to the single-shift case, \eqref{eq:approximate_multi_model_first_stage} needs to be solved using two-layer nested simulation.
In the outer layer, $M$ values of $\bLambda$ are sampled from the joint distribution of $\bLambda$, which are denoted as $\blambda_{m}$, $m=1,\ldots,M$.
In the inner layer, $g(\bk,\blambda,\pi_0)$ is estimated for given $\bk$ and $\blambda$ with $R$ independent simulation replications.
A convenient way to estimate $g(\bk,\blambda,\pi_0)$ is to generate the entire sample path from shift $1$ to $I$ with given $\bk$ and $\blambda$, for $R$ times.
Each sample path tracks the complete trajectory of the service system state over time, based on which a random sample of $g(\bk,\blambda,\pi_0)$ can be computed and denoted as $G_r(\bk,\blambda,\pi_0)$.
Then $g(\bk,\blambda,\pi_0)$ is estimated by
$$
\bar{g}_R (\bk,\blambda,\pi_0) \coloneqq \frac{1}{R} \sum_{r=1}^R G_r(\bk,\blambda,\pi_0).
$$
Thus, the optimization problem for deciding the optimal base staffing level $\bb$ in SOA is
\begin{equation}
\label{eq:simulated_multi_model_first_stage}
\bar{\tilde{\bb}}^* \coloneqq \arg\min_{\bb \in \NatInt^I}  \qty{ \sum_{i=1}^I c_i^{(b)} b_i   + \frac{1}{M} \sum_{m=1}^M \qty [ \min_{\bs\in\NatInt^I} \qty{ \sum_{i=1}^I c_i^{(s)} s_i + \bar{g}_R(\bb+\bs,\blambda_m,\pi_0) } ] }.
\end{equation}

To solve $\bar{\tilde{\bb}}^*$ in \eqref{eq:simulated_multi_model_first_stage}, we adopt the coordinate-search algorithm proposed by \citet{hongDiscreteOptimizationSimulation2005}, which is a coordinate-wise hill-climbing search algorithm.
Define
\[
    h_m(\bb,\bs)
    \coloneqq
    \sum_{i=1}^{I} c_i^{(s)}s_i
    +
    \bar{g}_R(\bb+\bs,\blambda_m,\pi_0),
\]
and $\bs_m(\bb) \coloneqq \arg\min_{\bs\in\NatInt^I} h_m(\bb,\bs)$.
The entire procedure is summarized in Algorithm~\ref{alg:joint_simulation_optimization}.
The implementation strategies to enhance computational efficiency are analogous to those used in the single-shift case, including caching $\bar{g}_R (\bk,\blambda,\pi_0)$ by $(\bk,\blambda)$, utilizing CRN when evaluating $\bar{g}_R (\bk,\blambda,\pi_0)$ for different values of $\bk$ under the same $\blambda$, and parallelizing for $\blambda_{m}$, $m=1,\ldots, M$.

\begin{algorithm}[!b]
\caption{Coordinate-Search Simulation Optimization Algorithm for Base Staffing}
\label{alg:joint_simulation_optimization}
\small
\begin{algorithmic}[1]
\Require
Arrival rate realizations $\{\blambda_m\}_{m=1}^{M}$, initial state $\pi_0$;
outer search round $K_{\text{out}}$, inner search round $K_{\text{in}}$;
initial solution $\bb^{(0)}$.
\Ensure
Base staffing solution $\hat{\bb}$, corresponding joint surge solutions $\{\bs_m(\hat{\bb})\}_{m=1}^{M}$.
\Statex \textbf{Main Function in Outer Layer}
\State $\bb \gets \bb^{(0)}$
\For{$k_{\text{out}}=1,\ldots,K_{\text{out}}$}
    \For{$i=1,\ldots,I$}
        \State \parbox[t]{\dimexpr\linewidth-\algorithmicindent}{Update $b_i$ while fixing other elements in $\bb$ by one-dimensional hill-climbing search, via calling\\ \Call{Evaluate}{$\bb$} to estimate the objective function for each $\bb$.}
    \EndFor
\EndFor
\State $\hat{\bb} \gets \bb$.
\State \Return $\hat{\bb}$ and $\{\bs_m(\hat{\bb})\}_{m=1}^{M}$.

\Statex \textbf{Objective Function Estimation in Inner Layer}
\Function{Evaluate}{$\bb$}
    \For {$m=1, \ldots, M$ \textbf{in parallel}}
        \State Initialize $\bs_m$.

        \For{$k_{\text{in}}=1,\ldots,K_{\text{in}}$}
            \For{$j=1,\ldots,I$}
                \State \parbox[t]{\dimexpr\linewidth-\algorithmicindent}{Update $s_{mj}$ while fixing other elements in $\bs_m$ by one-dimensional hill-climbing search on\\ $h_m(\bb,\bs_m)$.}
            \EndFor
        \EndFor

    \EndFor
    \State \Return  $\displaystyle \sum_{i=1}^{I} c_i^{(b)}b_i + \frac{1}{M}\sum_{m=1}^{M} h_m(\bb,\bs_m)$, and associated solutions $\{\bs_m\}_{m=1}^{M}$.
\EndFunction
\end{algorithmic}
\end{algorithm}

Strictly speaking, to estimate $\tilde{s}^*_i$ in \eqref{eq:tilde_s_i}, one needs to select $\blambda_m$'s from the $M$ generated samples of $\bLambda$ such that their $i$-th element exactly equals to $\lambda_i$, and then calculate the average of $s_i(\tilde{\bb}^*,\blambda,\pi_0)$ over these $\blambda_m$'s.
However, due to the continuity of $\blambda$, the probability of existence of such $\blambda_m$'s is zero.
Then one has to re-generate samples of $\bLambda$ conditional on $\Lambda_i=\lambda_i$ and compute $s_i(\tilde{\bb}^*,\blambda,\pi_0)$ for each of them.
This computation can only start upon observing $\lambda_i$, whose time delay will violate the requirement of (near-)real-time decision making.

We propose to further simplify the computation by utilizing the offline-generated samples and computed results.
Let $\hat{\bb}$ be the solved base staffing and $\bs_m(\hat{\bb})$ the associated surge staffing solutions if actual arrival rates are revealed simultaneously as $\blambda_m$, $m=1,\ldots, M$,
which are returned by Algorithm~\ref{alg:joint_simulation_optimization}.
Denote $\lambda_{mi}$ as the $i$-th element of $\blambda_{m}$.
Divide the possible range of arrival rate (i.e., $(0,\infty)$) into $N$ disjoint intervals based on the quantiles of $\{\lambda_{1i},\ldots,\lambda_{Mi}\}$.
More specifically, order $\{\lambda_{1i},\ldots,\lambda_{Mi}\}$ as $\lambda_{(1)i} \leq \lambda_{(2)i} \leq \cdots \leq \lambda_{(M)i}$, and then the upper bound of each interval is defined as
\[
    d_{in} \coloneqq \frac{\lambda_{(\lfloor nM/N \rfloor)i} +\lambda_{(\lfloor nM/N \rfloor + 1)i}}{2},\quad n=1,\ldots,N-1.
\]
The first interval is $\calB_{i1} \coloneqq \big(0, d_{i1}\big]$, the last interval is $\calB_{iN} \coloneqq  \big(d_{i,N-1}, \infty \big)$, and the intervals in between are $\calB_{in} \coloneqq \big(d_{i,n-1}, d_{i,n} \big]$, $n=2,\ldots,N-1$.
For the observed $\lambda_i$, suppose it falls in $\calB_{in}$, where $n\in\{1,\ldots,N\}$, then the solved optimal surge staffing level is
\begin{equation}\label{eq:simulated_multi_model_surge}
\hat{s}_{in} \coloneqq \textrm{round} \qty( \frac{N}{M} \sum_{m: \ \lambda_{mi} \in \calB_{in}} s_{mi}(\hat{\bb})).
\end{equation}

In the implementation, to balance the computation burden and accuracy, after obtaining $\hat{\bb}$ and $\{\bs_m(\hat{\bb})\}_{m=1}^M$ for some $M$,
we can generate additional samples of $\bLambda$ for better arrival rate interval construction and surge staffing level aggregation.
Say, for some $M'>M$, additionally generate $\blambda_m$, $m=M+1,\ldots,M'$ and compute the associated $\{\bs_m(\hat{\bb})\}_{m=M+1}^{M'}$ by calling \textsc{Evaluate} function in Algorithm~\ref{alg:joint_simulation_optimization}.
Then the above interval construction and solution aggregation are with respect to $\{\lambda_{mi}\}_{m=1}^{M'}$ and $\{\bs_m(\hat{\bb})\}_{m=1}^{M'}$.
Note that the aforementioned method is only one possibility to realize the offline simulation online application idea.
While more sophisticated methodologies such as surrogate model construction could certainly be employed,
we adopt it due to its simplicity and satisfactory performance (which will be shown in numerical results).

\subsubsection{Two-Stage Staffing Lookup Table}

To be more conveniently used in practice, the estimated optimal surge staffing level $\hat{s}_{in}$ can be pre-computed for the possible values of $\lambda_i$ before we actually observe it.
Namely, for each shift $i$, we only need to compute $\hat{s}_{in}$ for each interval $\calB_{in}$, $n=1,\ldots,N$, and record it.
With this, we can make a two-stage staffing lookup table far before the successive shifts start.
Table \ref{tab:lookup_table_sketch} shows a sketch of such a staffing lookup table.
In practice, for adjacent intervals of $\lambda_i$, they can be merged together if their surge staffing levels are the same.

\begin{table}[!h]
\centering
\caption{Sketch of a two-stage staffing lookup table based on SOA.}
\label{tab:lookup_table_sketch}
\small
\setlength{\tabcolsep}{6pt}
\renewcommand{\arraystretch}{1.25}
\begin{tabular}{l *{2}{r@{${}\,\mapsto\,{}$}l} c r@{${}\,\mapsto\,{}$}l}
\toprule
Shift $i$
  & \multicolumn{2}{c}{$1$}
  & \multicolumn{2}{c}{$2$}
  & $\cdots$
  & \multicolumn{2}{c}{$I$} \\
\midrule
Solved base staffing $\hat{b}_i$
  & \multicolumn{2}{c}{$\hat{b}_1$}
  & \multicolumn{2}{c}{$\hat{b}_2$}
  & $\cdots$
  & \multicolumn{2}{c}{$\hat{b}_I$} \\
\midrule
\multirow{4}{*}{\makecell[l]{Solved surge staffing $\hat{s}_i$,\\for different intervals of $\lambda_i$}}
  & $\calB_{11}$ & $\hat{s}_{11}$ & $\calB_{21}$ & $\hat{s}_{21}$ & $\cdots$ & $\calB_{I1}$ & $\hat{s}_{I1}$ \\
  & $\calB_{12}$ & $\hat{s}_{12}$ & $\calB_{22}$ & $\hat{s}_{22}$ & $\cdots$ & $\calB_{I2}$ & $\hat{s}_{I2}$ \\
  & \multicolumn{2}{c}{$\vdots$}  & \multicolumn{2}{c}{$\vdots$}  & $\ddots$ & \multicolumn{2}{c}{$\vdots$} \\
  & $\calB_{1N}$ & $\hat{s}_{1N}$ & $\calB_{2N}$ & $\hat{s}_{2N}$ & $\cdots$ & $\calB_{IN}$ & $\hat{s}_{IN}$ \\
\bottomrule
\end{tabular}
\end{table}

Since both the base staffing decision and surge staffing decision are computed well in advance,
the substantial time required for simulation and optimization exerts negligible impact on real-time decision-making.
Users only need to check the staffing lookup table to decide the base staffing levels for all shifts in the beginning and the surge staffing level of each shift $i$ upon observing $\lambda_i$ (simply by finding the corresponding interval in the table that covers $\lambda_i$).
The two layers of approximation on the model side (proposed in \Cref{sec:approximated_model}) and the compromise approach used in estimating $\tilde{s}^*_i$ (proposed in \Cref{sec:approximated_solution}) of course make the final solutions (thus the two-stage staffing lookup table) sub-optimal.
However, as will be demonstrated shortly in the numerical experiments, the quality of this SOA-based staffing policy closely approaches the true optimal solution and significantly outperforms existing approaches in the literature.

\subsection{Comparison with Existing Policies} \label{sec:comparison}

Note that \citet{bassambooCapacitySizingParameter2010} and \citet{huPredictionDrivenSurgePlanning2025} have proposed explicit staffing policies for more specific settings in EDs.
We first summarize their setting and assumptions, then present their policies, along with necessary discussion on how our SOA is applied in their settings.

Their basic models share the same setting and assumptions.
For shift $i$, patient arrivals are assumed to follow a DSPP with random arrival rate $\Lambda_i$, which has the following structure:
\begin{equation}
    \label{eq:Lambda}
    \Lambda_i \coloneqq \bar{\lambda}_i + X_i \bar{\lambda}_i^{\alpha},
\end{equation}
where $X_i \sim \calN(0,\sigma^2)$, $\alpha \in (0,1)$ is a constant, and $\bar{\lambda}_i>0$ is a constant for each $i$.
Let $F_i$ denote the distribution of $\Lambda_i$, and it is easy to see that $\E[\Lambda_i]=\bar{\lambda}_i$.
Note that the support of $\Lambda_i$ in \eqref{eq:Lambda} contains negative values.
In practice, only the positive samples of $\Lambda_i$ are retained (i.e., the truncated version of $\Lambda_i$ is actually used).
Conditional on $\Lambda_i = \lambda_i$, the arrival process is a homogeneous Poisson process, and the ED is assumed to operate as an $M/M/k + M$ queue.
The service and patience times are independent exponential random variables with means $1 / \mu$ and $1 / \theta$, respectively, and are independent of the arrival process.
For each individual in the queue, when the waiting time reaches the patience time, s/he will leave the system immediately.
The initial condition is left unspecified, as they both treat the service system as if it operates in steady state, whereas the underlying problem inherently spans a finite horizon with multiple shift transitions.

The considered QoS includes the queue length and abandonment rate.
To be precise, the cost function over $I$ consecutive shifts is defined as follows.
For notational convenience, normalize the shift length to one (by adjusting all the rates), thus the total length for $I$ consecutive shifts is $I$.
Let $Q(t;\bk,\blambda,\pi_0)$ denote the queue length at time $t$,
and $B(t;\bk,\blambda,\pi_0)$ the cumulative number of abandonments up to time $t$, where $t\in[0,I]$.
Then, the expected penalty related to the QoS of the service system is
\[
    g(\bk,\blambda,\pi_0) = h\,\mathbb{E}\left[ \int_0^I Q(t;\bk,\blambda,\pi_0)\,\mathrm{d}t \right] + a \,\mathbb{E}\left[B(I;\bk,\blambda,\pi_0)\right],
\]
where expectations are taken with respect to the queueing randomness, $h$ is the waiting cost per customer per unit time and $a$ is the cost per abandonment.
However, applying their formulations separately to each shift gives
\[
    g(\bk,\blambda) = (h + a \theta)\, \sum_{i=1}^I q(k_i,\lambda_i),
\]
where $q(k_i,\lambda_i)$ denotes the expectation of the steady-state queue length,
as they treat the system as if it operates in steady state for each shift.
This is certainly a simplifying approximation, which constitutes one of the reasons for the sub-optimal performance of their policies in practice.

Based on such assumptions of the stylized models of arrival and service system, and the steady-state approximation of the QoS,
\citet{bassambooCapacitySizingParameter2010} propose a single-stage newsvendor policy in light of a stochastic-fluid approximation, which is denoted as $\calP_{1,\textrm{NV}}$ in this paper.
Formally, $\calP_{1,\textrm{NV}}$ specifies that
\[
b_i^{1,NV} = \frac{1}{\mu} F_i^{-1}\left(1 - \frac{c_i^{(b)}}{h \mu/ \theta + a\mu} \right),
\]
and $s_i^{1,NV} = 0$.
Under the arrival-rate scaling and regularity assumptions, $\calP_{1,\textrm{NV}}$ is proved to have an optimality gap of $O(\bar{\lambda}_i^{1-\alpha})$ relative to the optimal single-stage staffing cost for $\alpha\in(1/2,1)$, as $\bar{\lambda}_i\to\infty$ \citep{bassambooCapacitySizingParameter2010}.
\citet{huPredictionDrivenSurgePlanning2025} propose a two-stage QED policy by combining the fluid-based staffing rule with a diffusion approximation to account for queueing variability, which is denoted as $\calP_{2,\textrm{QED}}$ in this paper.
Formally, for the regime $h\mu/\theta+a\mu>c_i^{(s)}>c_i^{(b)}$, $\calP_{2,\textrm{QED}}$ specifies that
\[
b_i^{2,QED} = \frac{1}{\mu} F_i^{-1}\left(1-\frac{c_i^{(b)}}{c_i^{(s)}}\right) + \eta_i^*\sqrt{\frac{\bar{\lambda}_i}{\mu}},
\]
and, after observing $\lambda_i$,
\[
s_i^{2,QED}(\lambda_i) = \left(\frac{\lambda_i}{\mu} + \eta_i^*\sqrt{\frac{\lambda_i}{\mu}} - b_i^{2,QED} \right)^+,
\]
where $(x)^+=\max\{x,0\}$, and the parameter $\eta_i^*$ is obtained by solving
\[
\eta_i^* = \arg\min_{\eta\in\mathbb{R}} \left\{c_i^{(s)}\eta + \left(\frac{h\mu}{\theta}+a\mu\right) \frac{\sqrt{\theta/\mu}\left[H\!\left(\eta\sqrt{\mu/\theta}\right) - \eta\sqrt{\mu/\theta}\right]}{1 + \sqrt{\theta/\mu}\, H\!\left(\eta\sqrt{\mu/\theta}\right)/H(-\eta)}\right\},
\]
where $H(x)=\phi(x)/\Phi(-x)$ is the standard normal hazard function, and $\phi$ and $\Phi$ denote the standard normal density and distribution functions, respectively.
Under certain assumptions, $\calP_{2,\textrm{QED}}$ is proved to have an optimality gap of $o(\sqrt{\bar{\lambda}_i})$ relative to the optimal two-stage staffing cost for $\alpha\in(1/2,1)$, as $\bar{\lambda}_i\to\infty$ \citep{huPredictionDrivenSurgePlanning2025}.
Both policies need to round their staffing decisions to integers when deployed in practice.
For the multi-shift case, \citet{huPredictionDrivenSurgePlanning2025} suggest some heuristic adjustments for compensating the transient-shift effect:
\[
    b_i^{\mathrm{adj}} = b_i^{2,QED}+\xi_1\Delta_i, \qquad s_i^{\mathrm{adj}} = s_i^{2,QED}(\lambda_i)+\xi_2D_i,
\]
where $\Delta_i$ is the difference in expected steady-state queue lengths between the preceding and current shifts, and $D_i$ measures the deviation of the initial queue length from its steady-state expectation under the current staffing.
The parameters $\xi_1$ and $\xi_2$ control the base and surge adjustments, respectively.

\section{Numerical Experiments}\label{s:experiments}

Before presenting the experimental setup, we must provide a supplementary description of the operational rules of the service system under study.
When the staffing level $k_i$ varies among shifts, the service system must have a clear rule to handle the ``customers'' in the ``servers'' if the number of servers is reduced in a subsequent shift.
Throughout this paper, we consider the same simple rule as in \citet{huPredictionDrivenSurgePlanning2025}.
When the number of servers is about to be reduced in the beginning of a shift,
all ``customers'' in the ``servers'' have to suspend their service and rejoin the front of the waiting line, where an earlier arrival time corresponds to a more advanced position in the queue.
When they re-enter the server, their service can resume from where it was interrupted.

It is worth emphasizing that how the system handles entities currently in service when the number of servers decreases is merely an operational policy.
While this rule dictates the low-level execution of the simulation for performance estimation, it remains completely independent of our proposed simulation-based optimization framework.
This issue has rarely been mentioned in existing studies, because they neither conduct a rigorous analytical derivation nor a strict simulation for the multi-shift case.

We use CRNs to compare candidate staffing decisions.
For each arrival rate realization, all candidates share the same arrival, service-time, and patience-time samples within each replication to reduce simulation noise in cost comparisons.
To avoid repeated simulation, we cache the estimated waiting and abandonment costs for each arrival rate and total staffing configuration.
With the initial state and random samples unchanged, different base-surge combinations with the same total staffing reuse these estimates, while their staffing costs are calculated separately.
All numerical experiments are implemented in Python on a personal computer equipped with an Intel Core i5-11400 processor (2.60 GHz) and 32 GB of RAM.
Source code is available at \url{https://github.com/shenhaihui/soa}.

\subsection{Single-Shift Case}

We consider two versions for our proposed SOA.
The first version is to directly solve $\bar{b}_i^*$ in \eqref{eq:simulated_model_first_stage} and $\bar{s}_i^*$ in \eqref{eq:simulated_model_surge_stage}, which is denoted as $\calP_{2,\textrm{SOA}}$-D.
Note that for the single-shift case, $\calP_{2,\textrm{SOA}}$-D closely approaches the true optimum (up to the errors in simulation and optimization), because no approximations are introduced into the original formulation.
The second version is to construct the two-stage staffing lookup table by solving $\bar{b}_i^*$ in \eqref{eq:simulated_model_first_stage} and $\hat{s}_{in}$ in \eqref{eq:simulated_multi_model_surge}, which is denoted as $\calP_{2,\textrm{SOA}}$.
For comparison, we also implement $\calP_{1,\textrm{NV}}$ and $\calP_{2,\textrm{QED}}$, which are discussed in \Cref{sec:comparison}.
Moreover, to better investigate the benefit of SOA, we also consider two variants.
The first one is called $\calP_{1,\textrm{SOA}}$, which solves $\bar{b}_i^*$ in \eqref{eq:simulated_model_first_stage} by restricting $s_i=0$.
It means that we only consider the base staffing in SOA.
The second one is called $\calP_{\textrm{QED} + \textrm{SOA}}$, where the base staffing decision is made via $\calP_{2,\textrm{QED}}$ and the surge staffing decision is made via $\calP_{2,\textrm{SOA}}$-D by directly solving $\bar{s}_i^*$ in \eqref{eq:simulated_model_surge_stage}.
It is used to individually investigate the benefit of SOA in the first-stage decision (when compared to $\calP_{2,\textrm{SOA}}$-D or $\calP_{2,\textrm{SOA}}$) and the benefit of SOA in the second-stage decision (when compared to $\calP_{2,\textrm{QED}}$).

\subsubsection{Special Problem Setting}

We first consider the special problem setting of nurse staffing in EDs in \citet{huPredictionDrivenSurgePlanning2025} (as presented in \Cref{sec:comparison}).
Since \citet{huPredictionDrivenSurgePlanning2025} treat the system as if it operates in steady state,
the initial condition is not clearly specified there.
We will try different initial conditions.

\textbf{Experiment 1.}
To begin with, we let the system start empty and idle.
Other problem parameters are exactly the same as in \citet{huPredictionDrivenSurgePlanning2025}.
That is, the parameters in the random arrival rate \eqref{eq:Lambda} are $\bar\lambda_i = 100$, $\alpha = 0.75$ and $\sigma = 1$.
The parameters for the queue are $\mu = 1$ and $\theta = 0.1$,
and the cost parameters are $h = 1.5$, $a = 3$, $c_i^{(b)} = 1$, and $c_i^{(s)} = 1.5$.
In the implementation of SOA, $M = M' = 1{,}000$, $R=100$, and $N = 20$.
In $\calP_{1,\textrm{SOA}}$, the search of base staffing solution starts from the solution of $\calP_{1,\textrm{NV}}$;
in $\calP_{2,\textrm{SOA}}$-D and $\calP_{2,\textrm{SOA}}$, the search of base staffing solution and surge staffing solution starts from the solutions of $\calP_{2,\textrm{QED}}$ (hereinafter the same).
For evaluation, we re-generate $1{,}000$ arrival rates and run 100 simulation replications of the system under each scenario, independently of the offline search.
For $\calP_{2,\textrm{SOA}}$-D, the surge staffing is solved online with another 100 simulation replications of the system.

Table~\ref{tab:experiment1_empty} shows the performance comparison among six policies, and Table~\ref{tab:experiment1_lookup} shows the constructed staffing lookup table of $\calP_{2,\textrm{SOA}}$. 
Several key findings are as follows.
(1)
The average total costs under $\calP_{2,\textrm{SOA}}$-D and $\calP_{2,\textrm{SOA}}$ are almost the same,
indicating that constructing the lookup table offline as a substitute of real-time simulation optimization for surge staffing decision incurs negligible optimality loss.
(2)
The average total cost under $\calP_{2,\textrm{SOA}}$ is significantly lower than that under $\calP_{1,\textrm{NV}}$ (by 61.8\%) and that under $\calP_{2,\textrm{QED}}$ (by 54.2\%).
(3)
Comparing $\calP_{1,\textrm{SOA}}$ to $\calP_{2,\textrm{SOA}}$-D or $\calP_{2,\textrm{SOA}}$, for this specific problem instance, the second-stage decision plays a marginal role because the corresponding surge staffing level is negligible compared to the base staffing level.
The underlying reason is that the variance of the random arrival rate here is relatively small, which will be further substantiated in subsequent experiments.
(4)
Nevertheless, the single-stage policy $\calP_{1,\textrm{NV}}$ performs the worst because it yields an excessive base staffing level.
(5)
$\calP_{2,\textrm{QED}}$ also yields an inflated base staffing level--albeit slightly lower than that of $\calP_{1,\textrm{NV}}$--but further introduces a substantial surge staffing level, pushing its total staffing level close to that of $\calP_{1,\textrm{NV}}$.
This is because they are derived by treating the system as operating in a steady state, whereas in reality the system starts empty.
Consequently, the actual resource requirement during the transient phase is substantially lower than expected.
Subsequent experiments will further substantiate this point.
(6)
Comparing $\calP_{\textrm{QED} + \textrm{SOA}}$ to $\calP_{2,\textrm{SOA}}$-D (or $\calP_{2,\textrm{SOA}}$) and $\calP_{2,\textrm{QED}}$, it is clear that both the first- and second-stage decisions in the SOA framework have unique and irreplaceable advantages.

\begin{table}[!t]
\centering
\caption{Performance comparison of six policies for Experiment 1.}
\label{tab:experiment1_empty}
\small
\setlength{\tabcolsep}{6pt}
\renewcommand{\arraystretch}{1.25}
\begin{tabular}{lcccccc}
\toprule
Metric
& $\calP_{2,\textrm{SOA}}$-D
& $\calP_{2,\textrm{SOA}}$
& $\calP_{1,\textrm{NV}}$
& $\calP_{2,\textrm{QED}}$
& $\calP_{1,\textrm{SOA}}$
& $\calP_{\textrm{QED+SOA}}$ \\
\midrule
Average total cost
& 57.290 & 57.288 & 150.000 & 125.218 & 57.340 & 94.192 \\
Base staffing $b_i$
& 38 & 38 & 150 & 94 & 39 & 94 \\
Average surge staffing $s_i$
& 1.148 & 1.249 & 0.000 & 20.812 & 0.000 & 0.000 \\
Offline search time (s)
& $2{,}115$ & $2{,}115$
& -- & -- & $1{,}569$ & -- \\
Online search time (s)
& $1{,}079$ & -- & -- & -- & -- & 391 \\
\bottomrule
\end{tabular}
\end{table}

\begin{table}[!t]
\centering
\caption{Two-stage staffing lookup table of $\calP_{2,\textrm{SOA}}$ for the single-shift case in Experiment 1.}
\label{tab:experiment1_lookup}
\small
\setlength{\tabcolsep}{6pt}
\renewcommand{\arraystretch}{1.25}
\begin{threeparttable}
\begin{tabular}{l r@{${}\,\mapsto\,{}$}l}
\toprule
Shift $i$ & \multicolumn{2}{c}{$1$} \\
\midrule
Solved base staffing $\hat{b}_i$ & \multicolumn{2}{c}{$38$} \\
\midrule
\multirow{6}{*}{\makecell[l]{Solved surge staffing $\hat{s}_i$,\\for different intervals of $\lambda_i$}}
  & $(0,122.295]$       & $0$  \\
  & $(122.295,127.224]$ & $1$  \\
  & $(127.224,133.160]$ & $2$  \\
  & $(133.160,143.488]$ & $4$  \\
  & $(143.488,155.819]$ & $8$  \\
  & $(155.819,\infty)$  & $14$ \\
\bottomrule
\end{tabular}
\begin{tablenotes}[para,flushleft]
\footnotesize
\textit{Note:} Adjacent intervals with the same surge staffing are merged.
\end{tablenotes}
\end{threeparttable}
\end{table}

\textbf{Experiment 2.}
We examine the number of patients remaining in the system at the end of the shift.
Across all policies and simulation replications, the average is approximately 67, with the 5th percentile at around 29 and the 95th percentile at around 110.
In this regard, we repeat Experiment 1 by initializing the system with 30, 60, 90, and 120 unfinished patients, respectively.
It is worth noting that the decisions under $\calP_{1,\textrm{NV}}$ and $\calP_{2,\textrm{QED}}$ are unaffected by these variations, as they do not depend on the initial state.
Hence, the base staffing decision in $\calP_{\textrm{QED} + \textrm{SOA}}$ also persists.

Table~\ref{tab:experiment2_initialization} shows the comparisons of costs and staffing decisions among six policies over different initial states.
Several key findings are as follows.
(1)
The performances of $\calP_{2,\textrm{SOA}}$-D and $\calP_{2,\textrm{SOA}}$ are still close and dominantly better than other policies.
As the initial number of patients increases, their associated base staffing levels
increase, and the total costs also increase.
(2)
For $\calP_{1,\textrm{NV}}$ and $\calP_{2,\textrm{QED}}$, the average total costs are relatively insensitive to initial conditions, as their massive staffing levels buffer the QoS against moderate changes in the initial number of patients.
(3)
Comparing $\calP_{2,\textrm{QED}}$ with $\calP_{2,\textrm{SOA}}$, the cost savings delivered by $\calP_{2,\textrm{SOA}}$ gradually shrink from 54.2\% to 10.5\% as the initial number of patients increases from 0 to 120, demonstrating increasing performance of $\calP_{2,\textrm{QED}}$.
This aligns with the fact that $\calP_{2,\textrm{QED}}$ is derived under the assumption of steady-state operation.
A relatively heavier initial workload accelerates the system's convergence toward steady state within a finite horizon,
which enhances the validity of the approximation in $\calP_{2,\textrm{QED}}$, resulting in better performance.
A similar trend can be observed for $\calP_{1,\textrm{NV}}$.
(4)
Other observations are the same as in Experiment 1.

\begin{table}[!b]
\centering
\caption{Cost and staffing comparisons of six policies for Experiment 2.}
\label{tab:experiment2_initialization}
\small
\setlength{\tabcolsep}{6pt}
\renewcommand{\arraystretch}{1.25}
\begin{tabular}{ccccccc}
\toprule
\multirow{2}{*}{Initialization}
& \multicolumn{6}{c}{Average total cost} \\
\cmidrule(lr){2-7}
& $\calP_{2,\textrm{SOA}}$-D
& $\calP_{2,\textrm{SOA}}$
& $\calP_{1,\textrm{NV}}$
& $\calP_{2,\textrm{QED}}$
& $\calP_{1,\textrm{SOA}}$
& $\calP_{\textrm{QED+SOA}}$ \\
\midrule
30
& 72.788 & 72.785 & 150.000 & 125.218 & 72.968 & 94.799 \\
60
& 88.777 & 88.770 & 150.002 & 125.222 & 89.229 & 97.094 \\
90
& 105.797 & 105.797 & 150.022 & 125.608 & 106.613 & 105.875 \\
120
& 125.163 & 125.166 & 150.160 & 139.902 & 126.063 & 131.723 \\
\midrule
\multirow{2}{*}{Initialization}
& \multicolumn{6}{c}{Staffing
$(b_i,\operatorname{Avg.}\,s_i)$} \\
\cmidrule(lr){2-7}
& $\calP_{2,\textrm{SOA}}$-D
& $\calP_{2,\textrm{SOA}}$
& $\calP_{1,\textrm{NV}}$
& $\calP_{2,\textrm{QED}}$
& $\calP_{1,\textrm{SOA}}$
& $\calP_{\textrm{QED+SOA}}$ \\
\midrule
30
& (55, 2.030) & (55, 2.130) & (150, 0.000)
& (94, 20.812) & (56, 0.000) & (94, 0.000) \\
60
& (73, 2.849) & (73, 2.884) & (150, 0.000)
& (94, 20.812) & (75, 0.000) & (94, 0.034) \\
90
& (92, 3.384) & (92, 3.488) & (150, 0.000)
& (94, 20.812) & (95, 0.000) & (94, 2.541) \\
120
& (115, 2.541) & (115, 2.546) & (150, 0.000)
& (94, 20.812) & (118, 0.000) & (94, 17.498) \\
\bottomrule
\end{tabular}
\end{table}

\textbf{Experiment 3.}
We also try some different values for the parameters in the arrival process and service system.
For two different initial states (60 and 120 unfinished patients, respectively),
we vary the holding cost $h \in \{1.5, 20\}$ with $a = 2h$,
and the perturbation variance $\sigma^2 \in \{1, 3.5\}$.
For the new parameter combinations that were not examined in Experiment 2, we rerun the experiments.
Table~\ref{tab:experiment3_sensitivity} shows the numerical results.
All the previous observations remain.
Moreover, we can see that for a given \(h\), an increase in \(\sigma ^{2}\) further amplifies the advantage of $\calP_{2,\textrm{SOA}}$ over $\calP_{1,\textrm{NV}}$ and $\calP_{2,\textrm{QED}}$.
For example, for 120 initial unfinished patients, $h=1.5$ and $\sigma^2=1$, the cost saving delivered by $\calP_{2,\textrm{SOA}}$ over $\calP_{2,\textrm{QED}}$ is 10.5\%, and it increases to 18.4\% when $\sigma^2=3.5$.
It reveals that $\calP_{2,\textrm{SOA}}$ can better cover the larger variance of arrival rates.
When $h=20$ and $\sigma^2=3.5$, the cost saving is even higher, which is 43.5\%.
The entire results demonstrate the high robustness of $\calP_{2,\textrm{SOA}}$ among different situations.
Moreover, comparing $\calP_{2,\textrm{SOA}}$ with $\calP_{1,\textrm{SOA}}$ also reveals a similar fact.
Note that in Experiment 1 we find the second-stage decision of $\calP_{2,\textrm{SOA}}$ plays a marginal role, where $\sigma^2=1$.
In this experiment, when $\sigma^2=1$, the costs of $\calP_{2,\textrm{SOA}}$ and $\calP_{1,\textrm{SOA}}$ are still very close.
When $\sigma^2=3.5$, the advantage of $\calP_{2,\textrm{SOA}}$ over $\calP_{1,\textrm{SOA}}$ may become clear.
For example,
for 60 initial unfinished patients, $h=20$ and $\sigma^2=3.5$, the cost saving delivered by $\calP_{2,\textrm{SOA}}$ is 18.6\%.

\begin{table}[!t]
\centering
\caption{Cost comparison of four policies for Experiment 3.}
\label{tab:experiment3_sensitivity}
\small
\setlength{\tabcolsep}{6pt}
\renewcommand{\arraystretch}{1.25}
\begin{tabular}{ccccccc}
\toprule
Initialization & $h$ & $\sigma^2$
& $\calP_{2,\textrm{SOA}}$
& $\calP_{1,\textrm{NV}}$
& $\calP_{2,\textrm{QED}}$
& $\calP_{1,\textrm{SOA}}$ \\
\midrule
\multirow{4}{*}{60}
& 1.5 & 1.0 & 88.770  & 150.002 & 125.222 & 89.229  \\
& 20  & 1.0 & 107.215 & 183.000 & 136.746 & 118.426 \\
& 1.5 & 3.5 & 96.774  & 194.004 & 143.737 & 98.557  \\
& 20  & 3.5 & 118.640 & 256.000 & 155.985 & 145.775 \\
\addlinespace
\multirow{4}{*}{120}
& 1.5 & 1.0 & 125.166 & 150.160 & 139.902 & 126.063 \\
& 20  & 1.0 & 137.720 & 183.006 & 197.404 & 147.794 \\
& 1.5 & 3.5 & 132.539 & 194.076 & 162.362 & 135.146 \\
& 20  & 3.5 & 148.401 & 256.000 & 262.536 & 174.845 \\
\bottomrule
\end{tabular}
\end{table}

\subsubsection{General Problem Setting}

\textbf{Experiment 4.}
We now consider a situation where the arrival process and service system depart from the stylized models assumed in \citet{bassambooCapacitySizingParameter2010} and \citet{huPredictionDrivenSurgePlanning2025}.
Note that substantial structural changes---such as drastically modifying the QoS metrics or the associated cost functions---would cause analytical policies like $\calP_{1,\textrm{NV}}$ and $\calP_{2,\textrm{QED}}$ to conceptually break down and lose their mathematical validity.
So here we only consider localized variations.
Suppose the interarrival times follow a Gamma distribution with mean $1/\lambda_i$ and squared coefficient of variation (SCV) 2, given that the rate is $\lambda_i$.
The service times follow a lognormal distribution with mean $1/\mu$ and SCV 2.
Moreover, there is balking behavior in the queue, i.e., an arrival leaves immediately when the number of waiting patients exceeds a threshold $B$, and no cost is counted.
Let $B=20$, and the values of other parameters remain the same as in Experiment 1.
We call this setting a non-stylized model and the setting in Experiment 1 a stylized model.

$\calP_{1,\textrm{NV}}$ and $\calP_{2,\textrm{QED}}$ will make the same staffing decisions as in Experiment 1, because they are forced to treat the system strictly as a stylized model, and the equivalent rates remain identical.
For SOA, the difference in the setting can be well captured in the simulation implementation.
Table~\ref{tab:experiment4_general} shows the numerical results for two different initial states (60 and 120 unfinished patients, respectively).
It can be seen that when the actual problem deviates from the stylized model assumed in \citet{bassambooCapacitySizingParameter2010} and \citet{huPredictionDrivenSurgePlanning2025},
the advantage of $\calP_{2,\textrm{SOA}}$ over $\calP_{1,\textrm{NV}}$ and $\calP_{2,\textrm{QED}}$ may further amplify.
For example, for 60 initial unfinished patients, the cost saving delivered by $\calP_{2,\textrm{SOA}}$ over $\calP_{2,\textrm{QED}}$ increases from 29.1\% to 35.7\% when the actual problem changes from stylized model to non-stylized model.
The reason is that analytical policies derived under stylized models, such as $\calP_{2,\textrm{QED}}$, can not fully capture the key characteristics of the problem.
It can be inferred that if we consider more substantial structural changes in the settings, the performance gap could be widened even further.

\begin{table}[!ht]
\centering
\caption{Cost and staffing comparisons of four policies for Experiment 4.}
\label{tab:experiment4_general}
\small
\setlength{\tabcolsep}{6pt}
\renewcommand{\arraystretch}{1.25}
\begin{tabular}{clcccc}
\toprule
\multirow{2}{*}{Initialization}
& \multirow{2}{*}{Model}
& \multicolumn{4}{c}{Average total cost} \\
\cmidrule(lr){3-6}
& &
$\calP_{2,\textrm{SOA}}$
& $\calP_{1,\textrm{NV}}$
& $\calP_{2,\textrm{QED}}$
& $\calP_{1,\textrm{SOA}}$ \\
\midrule
\multirow{2}{*}{60}
& Stylized
& 88.770 & 150.002 & 125.222 & 89.229 \\
& Non-stylized
& 80.570 & 150.002 & 125.223 & 80.544 \\
\addlinespace
\multirow{2}{*}{120}
& Stylized
& 125.166 & 150.160 & 139.902 & 126.063 \\
& Non-stylized
& 109.361 & 150.066 & 133.186 & 109.361 \\
\midrule
\multirow{2}{*}{Initialization}
& \multirow{2}{*}{Model}
& \multicolumn{4}{c}{Staffing
$(b_i,\operatorname{Avg.}\,s_i)$} \\
\cmidrule(lr){3-6}
& &
$\calP_{2,\textrm{SOA}}$
& $\calP_{1,\textrm{NV}}$
& $\calP_{2,\textrm{QED}}$
& $\calP_{1,\textrm{SOA}}$ \\
\midrule
\multirow{2}{*}{60}
& Stylized
& (73, 2.884) & (150, 0.000)
& (94, 20.812) & (75, 0.000) \\
& Non-stylized
& (62, 0.268) & (150, 0.000)
& (94, 20.812) & (62, 0.000) \\
\addlinespace
\multirow{2}{*}{120}
& Stylized
& (115, 2.546) & (150, 0.000)
& (94, 20.812) & (118, 0.000) \\
& Non-stylized
& (69, 0.000) & (150, 0.000)
& (94, 20.812) & (69, 0.000) \\
\bottomrule
\end{tabular}
\end{table}

\subsection{Multi-Shift Case}

As in \citet{huPredictionDrivenSurgePlanning2025}, for the multi-shift case, correlation between arrival rates in consecutive shifts is considered, where the $X_i$'s in \eqref{eq:Lambda} satisfy
\begin{equation}
\label{eq:correlation}
X_1 = \sigma Z_1, \qquad X_i = \rho X_{i - 1} + \sigma \sqrt{1 - \rho^2} Z_i, \quad i = 2, \ldots, I,
\end{equation}
where $Z_1, \cdots, Z_I$ are independent standard normal random variables, and $\rho$ controls the correlation between consecutive shifts.
In contrast to the single-shift case, the SOA framework for the general multi-shift case relies additionally on two layers of approximation on the model side.
We first try to investigate the ``optimality gap'' of $\calP_{2,\textrm{SOA}}$ in a very small-scale problem.
Then, on the case study of nurse staffing with real data investigated in \citet{huPredictionDrivenSurgePlanning2025}, $\calP_{2,\textrm{SOA}}$ is compared with other policies.

\subsubsection{Small Scale}

For very small problem size, it is possible to directly solve the original optimization problem, i.e., $\bb^*$ in \eqref{eq:original_multi_model_first_stage} and $s_i^*$ in \eqref{eq:original_multi_model_surge_decision}, using simulation optimization.
Such a solution is treated as the ``true optimal'' solution (of course subject to simulation optimization error), and we use $\calP_{2,\textrm{OPT}}$ to denote the corresponding policy.
We use $\calP_{2,\textrm{SOA}}$-D to denote the situation that $\bar{\tilde{\bb}}^*$ in \eqref{eq:simulated_multi_model_first_stage} is solved for base staffing, and
$\tilde{s}^*_i$ in \eqref{eq:tilde_s_i} is solved for surge staffing, which means we re-generate samples of $\bLambda$ conditional on $\Lambda_i=\lambda_i$ and do not use the offline-generated sample paths via the lookup table.
It is, of course, impossible for (near-)real-time decision in the general multi-shift case.
We use $\calP_{2,\textrm{SOA}}$-D to see the optimality gap due to the approximation on the model side.
The practical $\calP_{2,\textrm{SOA}}$ is also applied to investigate the overall optimality gap.

\textbf{Experiment 5.}
Consider the smallest multi-shift problem, i.e., $I=2$.
It is the simple extension of the problem in Experiment 1, with $\rho=0.3$ in \eqref{eq:correlation} representing a positive correlation, and all other parameters the same as in Experiment 1.
The system starts empty, and patients at the end of shift 1 carry over to shift 2.
In the implementation of $\calP_{2,\textrm{SOA}}$, $M=100$, $M'=1000$, $R=20$, $N=20$, $K_{\text{out}} = 1$  and $K_{\text{in}} = 2$.
For evaluation, we re-generate $100$ arrival rate vectors and run 100 simulation replications of the system under each scenario, independently of the offline search.
$\calP_{2,\textrm{OPT}}$ is implemented as a standard dynamic programming procedure, where the objective function evaluation is simulation-based, and the solution search is powered by the same coordinate-search algorithm as in $\calP_{2,\textrm{SOA}}$.
For $\calP_{2,\textrm{SOA}}$-D, to solve the surge staffing online, for each re-generated arrival rate vector used in evaluation, additional conditional arrival rate vectors are generated for the conditional expectation estimation in \eqref{eq:tilde_s_i}.
More details on the implementation can be found in the open-source code.

\begin{table}[!b]
\centering
\caption{Optimality gap and computation time for Experiment 5.}
\label{tab:two_shift_comparison}
\small
\setlength{\tabcolsep}{6pt}
\renewcommand{\arraystretch}{1.25}
\begin{threeparttable}
\begin{tabular}{lccccc}
\toprule
Policy
& \makecell{Total cost\\(95\% CI)}
& $(b_1,b_2)$
& $(\bar{s}_1,\bar{s}_2)$
& \makecell{Offline\\time (s)}
& \makecell{Online\\time (s)} \\
\midrule
$\calP_{2,\textrm{OPT}}$
& $159.233 \pm 8.073$
& $(42,78)$
& $(4.470,9.191)$
& $17{,}196$
& $11{,}063$ \\
$\calP_{2,\textrm{SOA}}$-D
& $164.643 \pm 8.563$
& $(42,79)$
& $(3.500,9.070)$
& $457$
& $724$ \\
$\calP_{2,\textrm{SOA}}$
& $164.135 \pm 8.607$
& $(42,79)$
& $(3.510,8.140)$
& $848$
& -- \\
\bottomrule
\end{tabular}
\begin{tablenotes}[para,flushleft]
\footnotesize
\textit{Note:} Offline times include base staffing search and, for $\calP_{2,\textrm{SOA}}$, lookup table construction.
\end{tablenotes}
\end{threeparttable}
\end{table}

\begin{table}[!b]
\centering
\caption{Two-stage staffing lookup table of $\calP_{2,\textrm{SOA}}$ for the two-shift case in Experiment 5.}
\label{tab:two_shift_lookup}
\small
\setlength{\tabcolsep}{6pt}
\renewcommand{\arraystretch}{1.25}
\begin{threeparttable}
\begin{tabular}{l *{2}{r@{${}\,\mapsto\,{}$}l}}
\toprule
Shift $i$
  & \multicolumn{2}{c}{$1$}
  & \multicolumn{2}{c}{$2$} \\
\midrule
Solved base staffing $\hat{b}_i$
  & \multicolumn{2}{c}{$42$}
  & \multicolumn{2}{c}{$79$} \\
\midrule
\multirow{15}{*}{\makecell[l]{Solved surge staffing $\hat{s}_i$,\\for different intervals of $\lambda_i$}}
  & $(0,112.370]$       & $0$  & $(0,74.362]$        & $0$  \\
  & $(112.370,117.478]$ & $1$  & $(74.362,79.221]$   & $1$  \\
  & $(117.478,121.664]$ & $3$  & $(79.221,83.183]$   & $2$  \\
  & $(121.664,126.947]$ & $4$  & $(83.183,87.802]$   & $3$  \\
  & $(126.947,132.199]$ & $6$  & $(87.802,95.596]$   & $4$  \\
  & $(132.199,138.871]$ & $9$  & $(95.596,103.520]$  & $5$  \\
  & $(138.871,153.504]$ & $13$ & $(103.520,106.868]$ & $6$  \\
  & $(153.504,\infty)$  & $24$ & $(106.868,110.478]$ & $7$  \\
  & \multicolumn{2}{c}{}       & $(110.478,115.249]$ & $8$  \\
  & \multicolumn{2}{c}{}       & $(115.249,120.393]$ & $6$  \\
  & \multicolumn{2}{c}{}       & $(120.393,125.745]$ & $14$ \\
  & \multicolumn{2}{c}{}       & $(125.745,131.740]$ & $16$ \\
  & \multicolumn{2}{c}{}       & $(131.740,138.190]$ & $17$ \\
  & \multicolumn{2}{c}{}       & $(138.190,149.118]$ & $19$ \\
  & \multicolumn{2}{c}{}       & $(149.118,\infty)$  & $28$ \\
\bottomrule
\end{tabular}
\begin{tablenotes}[para,flushleft]
\footnotesize
\textit{Note:} Adjacent intervals with the same surge staffing are merged.
\end{tablenotes}
\end{threeparttable}
\end{table}

Table~\ref{tab:two_shift_comparison} shows the optimality gap and computation time and Table~\ref{tab:two_shift_lookup} shows the constructed staffing lookup table of $\calP_{2,\textrm{SOA}}$. 
It can be seen that, at least for this two-shift problem, the two layers of model-side approximations are tight,
as the performance gap between $\calP_{2,\textrm{SOA}}$-D and the ``true optimal'' policy $\calP_{2,\textrm{OPT}}$ is only around 3\% in mean.
Furthermore, consistent with the findings in the single-shift case, constructing the lookup table offline as a substitute for real-time simulation-optimization for surge staffing decisions incurs a negligible loss of optimality,
as the performances of $\calP_{2,\textrm{SOA}}$-D and $\calP_{2,\textrm{SOA}}$ are indistinguishable.
Note that the computation time for $\calP_{2,\textrm{OPT}}$ is substantially longer than that for $\calP_{2,\textrm{SOA}}$, and $\calP_{2,\textrm{SOA}}$ effectively shifts the online computational
burden to offline.
Also note that in the constructed staffing lookup table, due to the existence of errors, the surge staffing does not strictly increase with realized arrival rates, which is not reasonable.
Two refinement directions exist: either appending additional offline simulations around anomalous results to improve local precision, or applying a rule-based post-processing step to enforce monotonicity.
We do not elaborate further on these methods in the present paper.

\subsubsection[Practical Case in Hu et al. (2025)]{Practical Case in \citet{huPredictionDrivenSurgePlanning2025}}

Now we consider the practical one-week ED staffing problem with real data, investigated in \citet{huPredictionDrivenSurgePlanning2025}.
It consists of $I = 14$ consecutive 12-hour shifts,
and for convenience the shift length is normalized to one.
The expected arrival rate for each shift, $\bar{\lambda}_i$, is as shown in Table~\ref{tab:case_shift_arrivals}.
The arrival process is assumed to satisfy the models \eqref{eq:Lambda} and \eqref{eq:correlation} with parameters $\alpha = 0.769, \sigma^2=0.118$ and $\rho=0.3$.
The ED is assumed to operate as an $M/M/k + M$ queue given the realized arrival rate, with $\mu = 1.4$ and $\theta = 12/27.5$.
The staffing cost parameters are $c_i^{(b)}=540$ and $c_i^{(s)}=810$ (corresponding to hourly wages of $\$45$ and $\$67.5$, respectively), and the waiting and abandonment cost parameters are $h = 810$ and $a = 1620$.
The system starts empty, and unfinished patients carry over across shifts.

\begin{table}[!b]
\centering
\caption{Shift-level mean arrival rates based on the ED data \citep{huPredictionDrivenSurgePlanning2025}.}
\label{tab:case_shift_arrivals}
\small
\renewcommand{\arraystretch}{1.25}
\begin{tabular}{lccccccc}
\toprule
 & Sun & Mon & Tue & Wed & Thu & Fri & Sat \\
\midrule
Day $\bar\lambda_i$ & 141.019 & 207.385 & 188.769 & 186.942 & 185.208 & 175.173 & 147.058 \\
Night $\bar\lambda_i$ & 89.462 & 97.058 & 97.769 & 93.711 & 95.189 & 96.692 & 94.115 \\
\bottomrule
\end{tabular}
\end{table}

We compare $\calP_{1,\textrm{NV}}$, $\calP_{2,\textrm{QED}}$, $\calP_{1,\textrm{SOA}}$, and $\calP_{2,\textrm{SOA}}$.
For $\calP_{2,\textrm{QED}}$, the suggested heuristic adjustments in \citet{huPredictionDrivenSurgePlanning2025} as described in Section~\ref{sec:comparison} are applied, where $\xi_1=5$ and $\xi_2=1$.
In the implementation of $\calP_{2,\textrm{SOA}}$, $M=100$, $M'=4000$, $R=10$, $N=20$, $K_{\text{out}} = 1$  and $K_{\text{in}} = 2$.
We have tried larger $K_{\text{out}}$ and found that the performance improvement is marginal (see Appendix~\hyperlink{app-A}{A} for details).
For evaluation, we re-generate $100$ arrival rate vectors and run 100 simulation replications of the system under each scenario, independently of the offline search.

\begin{figure}[!b]
\centering
\includegraphics[width=\textwidth]{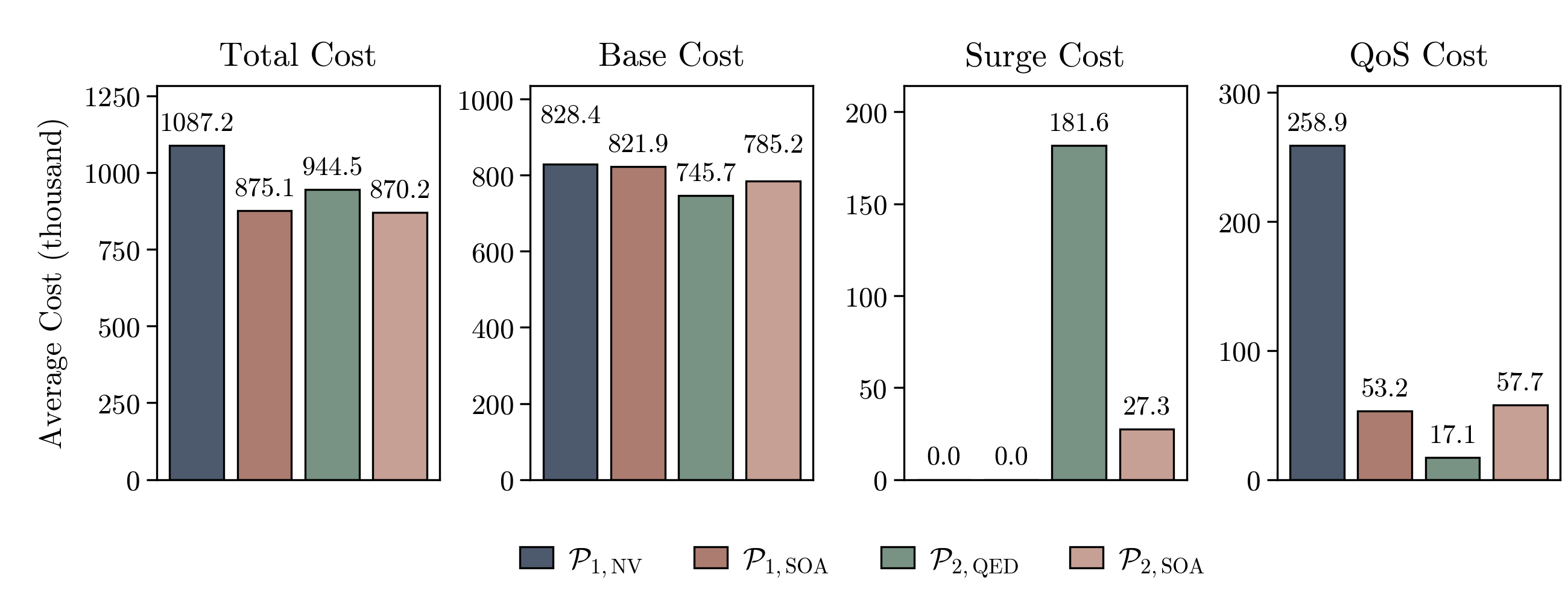}
\vspace{-0.8cm}
\caption{Comparison of average total cost and cost components under four policies.}
\label{fig:case_cost_decomposition}
\end{figure}

Table~\ref{tab:case_benchmark_summary} shows the performance of four policies, and
the full operational two-stage staffing lookup table constructed and used in $\calP_{2,\textrm{SOA}}$ is collected in Appendix~\hyperlink{app-B}{B}.
We can see that, in terms of the total cost, $\calP_{2,\textrm{SOA}}$ performs the best.
The cost saving delivered by $\calP_{2,\textrm{SOA}}$ is 19.96\% over $\calP_{1,\textrm{NV}}$ in mean, and
7.87\% over $\calP_{2,\textrm{QED}}$ in mean.
We can see that $\calP_{2,\textrm{QED}}$ yields an excessive surge staffing level, which makes its total staffing level the highest.
Although the resulting QoS is the best for all metrics, it does not trade off well with the staffing cost, which can be more clearly seen in Figure~\ref{fig:case_cost_decomposition}.

\begin{table}[!b]
\centering
\caption{Performance comparison of four policies for the nurse-staffing case.}
\label{tab:case_benchmark_summary}
\small
\setlength{\tabcolsep}{6pt}
\renewcommand{\arraystretch}{1.25}
\begin{tabular}{lcccccccc}
\toprule
Policy
& \makecell{Total cost\\(95\% CI)}
& \makecell{Offline\\time (s)}
& $\sum_i b_i$
& \makecell{Avg.\\$\sum_i s_i$}
& \makecell{Avg.\\queue}
& \makecell{Avg. wait\\(min)}
& \makecell{Wait $>10$\\min (\%)}
& \makecell{LWBS\\(\%)} \\
\midrule
$\calP_{1,\textrm{NV}}$
& $1{,}087{,}236 \pm 11{,}490$ & --
& 1534 & 0.00 & 14.086 & 74.72 & 43.68 & 3.220 \\
$\calP_{1,\textrm{SOA}}$
& $875{,}056 \pm 6{,}379$ & $1{,}691$
& 1522 & 0.00 & 2.570 & 13.43 & 18.36 & 0.768 \\
$\calP_{2,\textrm{QED}}$
& $944{,}474 \pm 7{,}889$ & --
& 1381 & 224.25 & 0.802 & 4.21 & 9.28 & 0.257 \\
$\calP_{2,\textrm{SOA}}$
& $870{,}178 \pm 7{,}207$ & $32{,}022$
& 1454 & 33.70 & 2.799 & 14.69 & 20.82 & 0.834 \\
\bottomrule
\end{tabular}
\end{table}

Another interesting thing is, the one-stage policy $\calP_{1,\textrm{SOA}}$ has very close cost performance compared with the two-stage policy $\calP_{2,\textrm{SOA}}$.
However, upon closer inspection, they differ in several fine details.
First, $\calP_{2,\textrm{SOA}}$ does decide some surge staffing levels, unlike $\calP_{1,\textrm{SOA}}$.
Second, although $\calP_{1,\textrm{SOA}}$ yields a higher total staffing level, the resulting QoS is relatively close to that in $\calP_{2,\textrm{SOA}}$.
Of course, one should note that QoS is not only determined by the total staffing level, but also by the detailed configuration among different shifts.
As a side note, $\calP_{1,\textrm{NV}}$ yields an even slightly higher total staffing level compared to $\calP_{1,\textrm{SOA}}$, but its resulting QoS is much worse.
Third, as one surge staff costs more than one base staff, in terms of the total cost, $\calP_{1,\textrm{SOA}}$ and $\calP_{2,\textrm{SOA}}$ exhibit no substantial difference.

\begin{table}[!b]
\centering
\caption{Comparison of average total cost with different demand variances and correlations.}
\label{tab:case_sensitivity}
\small
\setlength{\tabcolsep}{6pt}
\renewcommand{\arraystretch}{1.25}
\begin{tabular}{ccccccc}
\toprule
\multirow{2}{*}{$\sigma^2$}
& \multirow{2}{*}{$\rho$}
& \multicolumn{3}{c}{Average total cost}
& \multicolumn{2}{c}{Saving of $\calP_{2,\textrm{SOA}}$} \\
\cmidrule(lr){3-5} \cmidrule(lr){6-7}
& & $\calP_{1,\textrm{SOA}}$
& $\calP_{2,\textrm{QED}}$
& $\calP_{2,\textrm{SOA}}$
& vs.\ $\calP_{1,\textrm{SOA}}$
& vs.\ $\calP_{2,\textrm{QED}}$ \\
\midrule
0.118 & 0.0
& $870{,}197$ & $944{,}951$ & $868{,}040$
& 0.25\%
& 8.14\% \\
0.118 & 0.3
& $875{,}056$ & $944{,}474$ & $870{,}178$
& 0.56\%
& 7.87\% \\
\addlinespace
0.236 & 0.0
& $889{,}599$ & $960{,}977$ & $884{,}213$
& 0.61\%
& 7.99\% \\
0.236 & 0.3
& $898{,}385$ & $962{,}254$ & $888{,}149$
& 1.14\%
& 7.70\% \\
\addlinespace
0.354 & 0.0
& $907{,}179$ & $1{,}032{,}099$ & $898{,}680$
& 0.94\%
& 12.93\% \\
0.354 & 0.3
& $919{,}138$ & $1{,}033{,}761$ & $903{,}503$
& 1.70\%
& 12.60\% \\
\bottomrule
\end{tabular}
\end{table}

We further examine the policy performance under different levels of demand uncertainty and correlation between consecutive shifts.
Specifically, we consider $\sigma^2 \in \{0.118, 0.236, 0.354\}$ and $\rho \in \{0, 0.3\}$ and repeat the experiments, while keeping all other parameters unchanged.
Table~\ref{tab:case_sensitivity} reports the mean total costs of three policies and the cost savings achieved by $\calP_{2,\textrm{SOA}}$.
It can be seen that an increase in demand uncertainty (i.e., in $\sigma^2$) will further amplify the advantage of $\calP_{2,\textrm{SOA}}$ over $\calP_{2,\textrm{QED}}$.
Specifically, the cost saving delivered by $\calP_{2,\textrm{SOA}}$ over $\calP_{2,\textrm{QED}}$
increases from about 8\% to about 13\% when $\sigma^2$ increases from 0.118 to 0.354.
This effect and its underlying reason are aligned with those in the single-shift case considered in Experiment 3.
However, surprisingly, comparing $\calP_{2,\textrm{SOA}}$ with $\calP_{1,\textrm{SOA}}$, the effect is not that strong, and the cost saving delivered by $\calP_{2,\textrm{SOA}}$ over $\calP_{1,\textrm{SOA}}$ increases slowly with $\sigma^2$.
Since this case study is based on real data, this observation reveals that in real applications the two-stage staffing framework may not always exhibit a pronounced advantage relative to the one-stage alternative, especially considering its added complexity in computation and execution.
Recall that in the single-shift case considered in Experiment 3, we do see situations where $\calP_{2,\textrm{SOA}}$ substantially outperforms $\calP_{1,\textrm{SOA}}$.
Thus, we are not concluding that the two-stage staffing framework is universally redundant;
rather, we offer a caveat that this more sophisticated staffing approach may not always be necessary in practice.

Finally, we consider the most realistic scenario, where an ED actually operates continuously week after week without resetting, although staffing schedules are planned weekly.
Even over a time horizon free of significant seasonality, meaning that the mean arrival rate profile remains identical across weeks, unfinished patients at the end of one week carry over to the next week and the demand correlations persist across consecutive weeks.
Admittedly, $\calP_{2,\textrm{SOA}}$ is pre-computed offline under the one-week setting, and the two-stage staffing lookup table depends solely on the distributions of $\Lambda_i$ and the specified initial state at the beginning of the week, whereas real-time states are not directly accounted for in surge staffing decisions.
However, the initial state at the beginning of each week is a consequence of previous operations and cannot be known a priori.
By assuming the same initial states 
and ignoring the demand correlations across consecutive weeks, the two-stage staffing lookup table of $\calP_{2,\textrm{SOA}}$ remains identical across weeks with the same mean arrival rate profiles.
A natural question is whether applying such a one-week policy cyclically in a long horizon remains effective.

To answer this question, we consider a continuous 52-week ($728$-shift) horizon under the baseline parameter setting ($\sigma^2 = 0.118$, $\rho = 0.3$), assuming that the arrival-rate distribution remains identical across weeks (whose means are as shown in Table~\ref{tab:case_shift_arrivals}) and the demand correlation (as modeled in \eqref{eq:correlation}) evolves continuously across weeks.
As previously noted, because unfinished patients carry over, each week is unlikely to start empty.
To accommodate this issue, we use the average number of unfinished patients remaining at the end of shift 14 (which is 75), obtained from the previous baseline one-week experiment under the benchmark policy $\calP_{2,\textrm{QED}}$, as the specified initial state for $\calP_{1,\textrm{SOA}}$ and $\calP_{2,\textrm{SOA}}$ under the one-week setting, and re-solve these two policies accordingly.
We generate $1{,}000$ independent 52-week arrival rate trajectories to evaluate the performance of the four policies.
Table~\ref{tab:case_rolling_52weeks} shows the weekly average performance.
Note that the weekly costs of all four policies are slightly higher than those in Table~\ref{tab:case_benchmark_summary}, because each week now absorbs carried-over workload rather than starting empty.
But the superiority of $\calP_{2,\textrm{SOA}}$ over $\calP_{1,\textrm{NV}}$ and $\calP_{2,\textrm{QED}}$ remains similar as before, and the advantage of $\calP_{2,\textrm{SOA}}$ over $\calP_{1,\textrm{SOA}}$ is still marginal.

\begin{table}[!t]
\centering
\caption{Weekly average performance of four policies over a continuous 52-week horizon.}
\label{tab:case_rolling_52weeks}
\small
\setlength{\tabcolsep}{6pt}
\renewcommand{\arraystretch}{1.25}
\begin{tabular}{lccccccc}
\toprule
Policy
& \makecell{Weekly cost\\(95\% CI)}
& \makecell{Weekly\\$\sum_i b_i$}
& \makecell{Avg. weekly\\$\sum_i s_i$}
& \makecell{Avg.\\queue}
& \makecell{Avg. wait\\(min)}
& \makecell{Wait $>10$\\min (\%)}
& \makecell{LWBS\\(\%)} \\
\midrule
$\calP_{1,\textrm{NV}}$
& $1{,}118{,}401 \pm 739$
& 1534 & 0.00 & 15.782 & 83.84 & 47.81 & 3.616 \\
$\calP_{1,\textrm{SOA}}$
& $905{,}817 \pm 445$
& 1564 & 0.00 & 2.970 & 15.77 & 20.72 & 0.897 \\
$\calP_{2,\textrm{QED}}$
& $965{,}666 \pm 485$
& 1381 & 246.83 & 0.940 & 4.99 & 10.93 & 0.304 \\
$\calP_{2,\textrm{SOA}}$
& $902{,}282 \pm 477$
& 1495 & 36.44 & 3.180 & 16.89 & 23.08 & 0.957 \\
\bottomrule
\end{tabular}
\end{table}

\section{Conclusion}\label{s:conclusion}

This paper proposes the SOA to solve the two-stage staffing problems with random arrival rates, aiming at more general arrival processes and service systems.
Applying simulation optimization to such a problem is far from trivial and inherently accompanied by critical challenges.
We effectively address these challenges by recasting the original formulation, designing an efficient coordinate-search algorithm, and constructing a two-stage staffing lookup table offline.
Comprehensive numerical experiments demonstrate that the proposed SOA has significantly better staffing performance and substantially broader applicability compared to existing approaches.
Promising future research directions include explicitly integrating pre-shift forecasting errors (albeit more accurate than long-term forecasts) into the simulation optimization framework, and refining the two-stage staffing lookup table to smooth out numerical anomalies and enforce strict monotonicity.


\bibliographystyle{chicago}
\bibliography{references}

\newpage
\appendix

\pdfbookmark[1]{Appendix}{link-app}
\hypertarget{app}{
\section*{Appendix}
}

\pdfbookmark[2]{A. Test of the Convergence of the Coordinate Search}{link-app-A}
\hypertarget{app-A}{
\subsection*{A. Test of the Convergence of the Coordinate Search}
}

\begin{figure}[!h]
    \centering
    \includegraphics[width=0.75\textwidth]{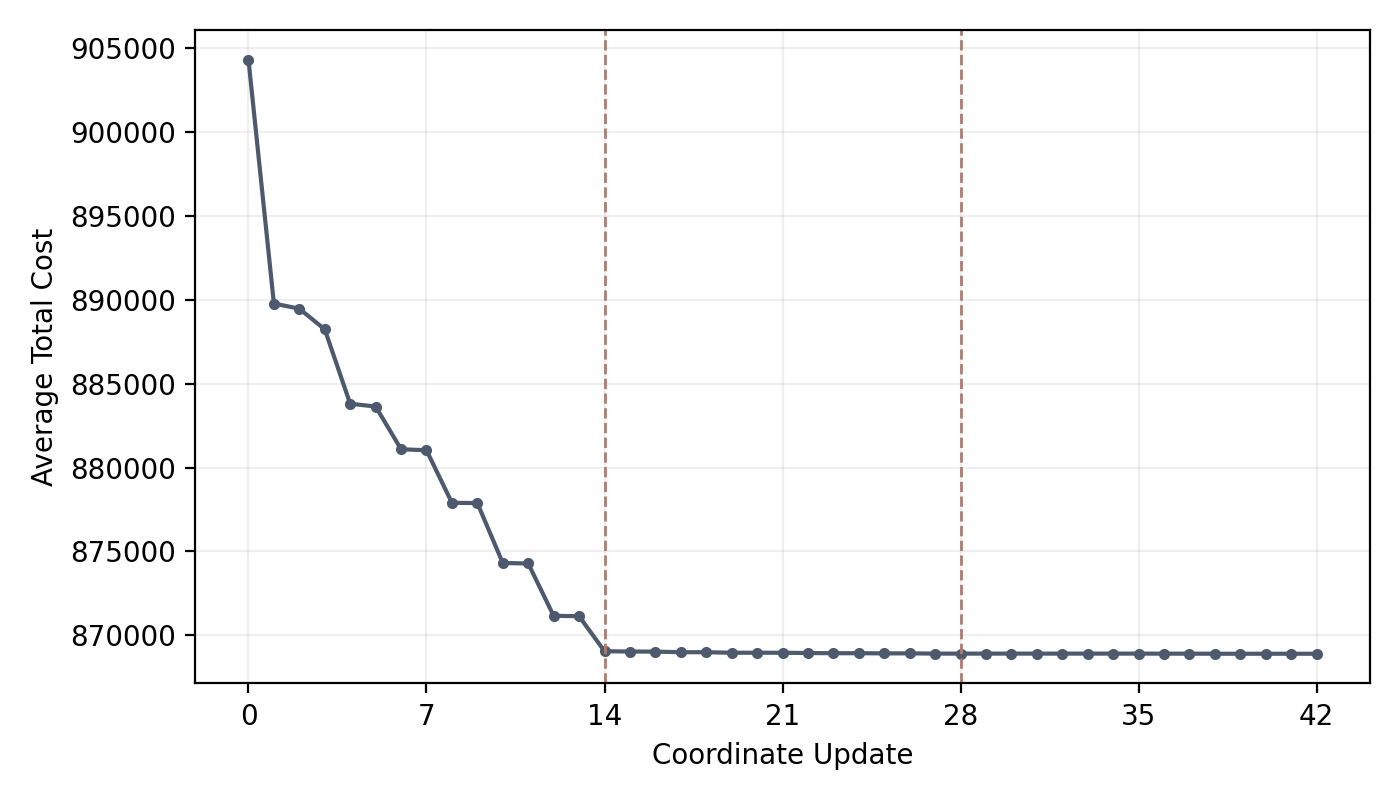}
    \caption{Progress of the outer coordinate search.
    (Dashed vertical lines distinguish different outer search rounds, each covering all $I=14$ coordinate searches.)}
    \label{fig:coordinate_gain}
\end{figure}

For the baseline one-week case, we examine the convergence of the coordinate-search algorithm when solving the base staffing level $\bar{\tilde{\bb}}^*$ in \eqref{eq:simulated_multi_model_first_stage}.
We set $K_{\text{out}} = 3$ and record the estimated objective function value in \eqref{eq:simulated_multi_model_first_stage} after each coordinate update.
in Figure~\ref{fig:coordinate_gain}.
It can be seen that the estimated objective function value tends to converge after the first round of outer search.

\pdfbookmark[2]{B. Full Staffing Lookup Table for the Nurse-Staffing Case}{link-app-B}
\hypertarget{app-B}{
\subsection*{B. Full Staffing Lookup Table for the Nurse-Staffing Case}
}

Table~\ref{tab:case_staffing_lookup} presents the full operational two-stage staffing lookup table of $\calP_{2,\textrm{SOA}}$ for the one-week (14-shift) nurse-staffing case.

\begin{sidewaystable}[!h]
\centering
\caption{Operational two-stage staffing lookup table of $\calP_{2,\textrm{SOA}}$ for the one-week (14-shift) nurse-staffing case.}
\label{tab:case_staffing_lookup}
\setlength{\tabcolsep}{1.5pt}
\renewcommand{\arraystretch}{1.05}
\resizebox{\linewidth}{!}{%
\begin{tabular}{@{}l *{7}{r@{${}\,\mapsto\,{}$}l}@{}}
\toprule
 & \multicolumn{2}{c}{\textbf{Sun}} & \multicolumn{2}{c}{\textbf{Mon}} & \multicolumn{2}{c}{\textbf{Tue}} & \multicolumn{2}{c}{\textbf{Wed}} & \multicolumn{2}{c}{\textbf{Thu}} & \multicolumn{2}{c}{\textbf{Fri}} & \multicolumn{2}{c}{\textbf{Sat}} \\
\midrule
\multicolumn{15}{@{}l}{\textit{Day shifts}} \\[2pt]
Shift $i$ & \multicolumn{2}{c}{1} & \multicolumn{2}{c}{3} & \multicolumn{2}{c}{5} & \multicolumn{2}{c}{7} & \multicolumn{2}{c}{9} & \multicolumn{2}{c}{11} & \multicolumn{2}{c}{13} \\
\makecell[l]{Solved base\\staffing $\hat{b}_i$} & \multicolumn{2}{c}{62} & \multicolumn{2}{c}{120} & \multicolumn{2}{c}{122} & \multicolumn{2}{c}{119} & \multicolumn{2}{c}{116} & \multicolumn{2}{c}{112} & \multicolumn{2}{c}{100} \\
\addlinespace[3pt]
\multirow{11}{*}{\makecell[l]{Solved surge\\staffing $\hat{s}_i$,\\by $\lambda_i$ interval}} & $(0,135.019]$ & $0$ & $(0,199.002]$ & $0$ & $(0,186.607]$ & $0$ & $(0,184.814]$ & $0$ & $(0,177.942]$ & $0$ & $(0,165.586]$ & $0$ & $(0,138.496]$ & $0$ \\
 & $(135.019,142.900]$ & $1$ & $(199.002,210.207]$ & $1$ & $(186.607,191.191]$ & $1$ & $(184.814,187.140]$ & $1$ & $(177.942,185.121]$ & $1$ & $(165.586,172.978]$ & $1$ & $(138.496,146.780]$ & $1$ \\
 & $(142.900,144.904]$ & $2$ & $(210.207,212.805]$ & $2$ & $(191.191,198.959]$ & $2$ & $(187.140,194.710]$ & $2$ & $(185.121,187.744]$ & $2$ & $(172.978,179.743]$ & $2$ & $(146.780,151.239]$ & $2$ \\
 & $(144.904,146.929]$ & $3$ & $(212.805,215.466]$ & $3$ & $(198.959,201.457]$ & $3$ & $(194.710,197.403]$ & $3$ & $(187.744,192.911]$ & $3$ & $(179.743,182.370]$ & $3$ & $(151.239,155.690]$ & $3$ \\
 & $(146.929,151.408]$ & $4$ & $(215.466,218.321]$ & $5$ & $(201.457,204.925]$ & $4$ & $(197.403,199.947]$ & $4$ & $(192.911,195.534]$ & $4$ & $(182.370,184.773]$ & $4$ & $(155.690,160.564]$ & $5$ \\
 & $(151.408,154.005]$ & $6$ & $(218.321,221.336]$ & $6$ & $(204.925,208.524]$ & $6$ & $(199.947,203.366]$ & $6$ & $(195.534,198.183]$ & $5$ & $(184.773,187.415]$ & $5$ & $(160.564,163.641]$ & $6$ \\
 & $(154.005,157.043]$ & $7$ & $(221.336,224.767]$ & $7$ & $(208.524,213.483]$ & $7$ & $(203.366,211.109]$ & $8$ & $(198.183,201.727]$ & $7$ & $(187.415,190.497]$ & $6$ & $(163.641,167.690]$ & $8$ \\
 & $(157.043,160.788]$ & $9$ & $(224.767,228.902]$ & $9$ & $(213.483,220.963]$ & $11$ & $(211.109,218.560]$ & $12$ & $(201.727,205.310]$ & $8$ & $(190.497,194.018]$ & $8$ & $(167.690,173.260]$ & $10$ \\
 & $(160.788,166.384]$ & $11$ & $(228.902,233.993]$ & $12$ & $(220.963,\infty)$ & $17$ & $(218.560,\infty)$ & $18$ & $(205.310,209.583]$ & $11$ & $(194.018,198.413]$ & $10$ & $(173.260,\infty)$ & $14$ \\
 & $(166.384,\infty)$ & $16$ & $(233.993,241.307]$ & $15$ & \multicolumn{2}{c}{} & \multicolumn{2}{c}{} & $(209.583,216.139]$ & $13$ & $(198.413,205.443]$ & $13$ & \multicolumn{2}{c}{} \\
 & \multicolumn{2}{c}{} & $(241.307,\infty)$ & $22$ & \multicolumn{2}{c}{} & \multicolumn{2}{c}{} & $(216.139,\infty)$ & $18$ & $(205.443,\infty)$ & $18$ & \multicolumn{2}{c}{} \\
\midrule
\multicolumn{15}{@{}l}{\textit{Night shifts}} \\[2pt]
Shift $i$ & \multicolumn{2}{c}{2} & \multicolumn{2}{c}{4} & \multicolumn{2}{c}{6} & \multicolumn{2}{c}{8} & \multicolumn{2}{c}{10} & \multicolumn{2}{c}{12} & \multicolumn{2}{c}{14} \\
\makecell[l]{Solved base\\staffing $\hat{b}_i$} & \multicolumn{2}{c}{76} & \multicolumn{2}{c}{113} & \multicolumn{2}{c}{108} & \multicolumn{2}{c}{107} & \multicolumn{2}{c}{106} & \multicolumn{2}{c}{103} & \multicolumn{2}{c}{90} \\
\addlinespace[3pt]
\multirow{9}{*}{\makecell[l]{Solved surge\\staffing $\hat{s}_i$,\\by $\lambda_i$ interval}} & $(0,81.976]$ & $0$ & $(0,94.299]$ & $0$ & $(0,91.628]$ & $0$ & $(0,84.188]$ & $0$ & $(0,90.449]$ & $0$ & $(0,90.711]$ & $0$ & $(0,88.177]$ & $0$ \\
 & $(81.976,90.677]$ & $1$ & $(94.299,102.992]$ & $1$ & $(91.628,99.425]$ & $1$ & $(84.188,85.970]$ & $1$ & $(90.449,99.760]$ & $1$ & $(90.711,92.392]$ & $1$ & $(88.177,95.579]$ & $1$ \\
 & $(90.677,95.124]$ & $2$ & $(102.992,112.195]$ & $2$ & $(99.425,102.301]$ & $2$ & $(85.970,87.779]$ & $0$ & $(99.760,103.037]$ & $2$ & $(92.392,93.998]$ & $0$ & $(95.579,100.057]$ & $2$ \\
 & $(95.124,96.785]$ & $3$ & $(112.195,116.117]$ & $4$ & $(102.301,103.775]$ & $1$ & $(87.779,99.558]$ & $1$ & $(103.037,110.052]$ & $3$ & $(93.998,99.644]$ & $1$ & $(100.057,103.613]$ & $3$ \\
 & $(96.785,100.885]$ & $4$ & $(116.117,\infty)$ & $6$ & $(103.775,107.425]$ & $2$ & $(99.558,103.245]$ & $2$ & $(110.052,114.404]$ & $5$ & $(99.644,102.917]$ & $2$ & $(103.613,108.638]$ & $4$ \\
 & $(100.885,103.391]$ & $5$ & \multicolumn{2}{c}{} & $(107.425,109.837]$ & $3$ & $(103.245,108.291]$ & $3$ & $(114.404,\infty)$ & $7$ & $(102.917,108.855]$ & $3$ & $(108.638,112.554]$ & $5$ \\
 & $(103.391,107.102]$ & $6$ & \multicolumn{2}{c}{} & $(109.837,117.071]$ & $4$ & $(108.291,112.459]$ & $4$ & \multicolumn{2}{c}{} & $(108.855,111.425]$ & $4$ & $(112.554,\infty)$ & $8$ \\
 & $(107.102,\infty)$ & $9$ & \multicolumn{2}{c}{} & $(117.071,\infty)$ & $8$ & $(112.459,\infty)$ & $7$ & \multicolumn{2}{c}{} & $(111.425,115.391]$ & $5$ & \multicolumn{2}{c}{} \\
 & \multicolumn{2}{c}{} & \multicolumn{2}{c}{} & \multicolumn{2}{c}{} & \multicolumn{2}{c}{} & \multicolumn{2}{c}{} & $(115.391,\infty)$ & $8$ & \multicolumn{2}{c}{} \\
\bottomrule
\end{tabular}}
\begin{flushleft}
\footnotesize
\textit{Note:}
Adjacent intervals with the same surge staffing are merged. Due to the existence of errors, the surge staffing does not strictly increase with realized arrival rates.
\end{flushleft}
\end{sidewaystable}

\end{document}